\documentclass[lettersize,journal]{IEEEtran}

\usepackage{amsmath,amsfonts}
\usepackage{algorithm}
\usepackage{algpseudocode} 

\usepackage{array}
\usepackage{textcomp}
\usepackage{stfloats}
\usepackage{url}
\usepackage{verbatim}
\usepackage{graphicx}
\usepackage{cite}

\usepackage[table,xcdraw]{xcolor}
\definecolor{customgreen}{HTML}{c0e9d2} 

\usepackage{cleveref}
\usepackage{siunitx}
\usepackage[inline]{enumitem}
\usepackage{booktabs}
\usepackage[table]{xcolor}
\usepackage{pifont}
\usepackage{caption}
\usepackage{subcaption}
\newcommand{\cmark}{\ding{51}}%
\newcommand{\xmark}{\ding{55}}%

\begin{document}

\title{ElectraSight: Smart Glasses with Fully Onboard Non-Invasive Eye Tracking Using Hybrid Contact and Contactless EOG} 

\author{Nicolas~Schärer, Federico~Villani, Aishwarya~Melatur, Steven~Peter, Tommaso~Polonelli, and Michele~Magno
\thanks{N. Schärer, F. Villani, A. Melatur, S. Peter, T. Polonelli and M. Magno are with the Center for Project-Based Learning of ETH Z\"urich, ETZ, Gloriastrasse 35, 8092 Z\"urich, Switzerland (e-mail: nicolas.schaerer@pbl.ee.ethz.ch, villanif@ethz.ch, aishwarya.melatur@pbl.ee.ethz.ch, steven.peter@alumni.ethz.ch, tommaso.polonelli@pbl.ee.ethz.ch, michele.magno@pbl.ee.ethz.ch).}%
\thanks{Special thanks to Alan Magdaleno, and all the anonymous people involved in the datasat collection. This work is supported by STMicroelectronics, which provided the sensors and technical support.}
}



\maketitle

\begin{abstract}
Smart glasses with integrated eye tracking technology are revolutionizing diverse fields, from immersive augmented reality experiences to cutting-edge health monitoring solutions. However, traditional eye tracking systems rely heavily on cameras and significant computational power, leading to high-energy demand and privacy issues. Alternatively, systems based on electrooculography (EOG) provide superior battery life but are less accurate and primarily effective for detecting blinks, while being highly invasive. The paper introduces ElectraSight, a non-invasive plug-and-play low-power eye tracking system for smart glasses. The hardware-software co-design of the system is detailed, along with the integration of a hybrid EOG (hEOG) solution that incorporates both contact and contactless electrodes. Within \qty{79}{\kilo\byte} of memory, the proposed tinyML model performs real-time eye movement classification with 81\% accuracy for 10 classes and 92\% for 6 classes, not requiring any calibration or user-specific fine-tuning. Experimental results demonstrate that ElectraSight delivers high accuracy in eye movement and blink classification, with minimal overall movement detection latency (90\% within \qty{60}{\milli\second}) and an ultra-low computing time (\qty{301}{\micro\second}). The power consumption settles down to \qty{7.75}{\milli\watt} for continuous data acquisition and \qty{46}{\milli\joule} for the tinyML inference. This efficiency enables continuous operation for over \qty{3}{days} on a compact \qty{175}{\milli\ampere\hour} battery. This work opens new possibilities for eye tracking in commercial applications, offering an unobtrusive solution that enables advancements in user interfaces, health diagnostics, and hands-free control systems.
\end{abstract}


\section{Introduction}
\label{sec:intro}

Eye tracking provides valuable insights into human visual behavior, medical diagnostic~\cite{yin_internet_2023}, attentional processes, and decision-making~\cite{yang_aim_2024} by decoding the movement of the eyes, gaze points, and blink patterns~\cite{klaib_eye_2021}. This technology has found extensive applications across various fields, including medicine~\cite{rashid_stress_2023}, consumer market~\cite{wang_airmouse_2019, kuang_deyeauth_2024}, and engineering~\cite{gutierrez_framework_2021}. In the medical domain, eye tracking is utilized in cognitive rehabilitation and as an assistive tool for patients with neurodegenerative conditions such as amyotrophic lateral sclerosis (ALS)~\cite{chen_iot_2022} or facial paralysis~\cite{cervera-negueruela_bionic_2024}, exemplified by its use in aiding renowned physicist Stephen Hawking~\cite{rezvani_review_2024}. In commercial contexts, eye tracking is used to assess consumer preferences and optimize product designs. In engineering, it plays a key role in enhancing human-computer interactions, especially in virtual reality (VR) systems~\cite{kang_real-time_2021}. 
The rapid growth of the wearable electronics market is driving significant advancements in human-machine interaction~\cite{polonelli_h-watch_2021}, with smart glasses steadily gaining adoption. VR, augmented reality (AR), and wearable devices are no longer limited to entertainment or fitness but are reshaping the concept of traditional glasses. What was once a purely optical tool is evolving into a sophisticated portable computing device, equipped with integrated sensors and intelligent features, offering new possibilities for interaction and functionality~\cite{li_humancomputer_2022}.

Given its broad utility, the development and refinement of eye tracking systems are crucial for advancing eye-based health monitoring~\cite{klaib_eye_2021}, improving human-machine interfaces~\cite{li_humancomputer_2022}, and enabling in-depth commercial analysis~\cite{zhang_online_2022}. Various methodologies have been proposed in the literature for eye tracking, including video oculography, scleral search coils, magnetic resonance-based systems, and invasive electrooculography (EOG). However, each of these techniques faces limitations that prevent their adoption outside of research\cite{klaib_eye_2021, gao_toward_2022}. For example, the scleral search coil method is invasive~\cite{li_review_2021}, requiring the placement of a coil in the eye, which can lead to discomfort, infection risks, and slippage as the eye moves. Magnetic resonance-based systems, while non-invasive, are dependent on large and cumbersome equipment, which compromises their portability and practical application in everyday settings~\cite{li_review_2021}. Also, standard EOG, which is one of the standard approaches for eye tracking, needs contact—often wet—electrodes that need to be placed in key positions~\cite{alam_high_2021}. Despite being the optimal solution so far, it is not adequate for \textit{plug-and-play} consumer devices such as smart glasses~\cite{frey_gapses_2024, kang_real-time_2021,moosmann_ultra-efficient_2023}. 

Another popular method relies on visual-based sensing, which proves effective and scalable across various human subjects~\cite{bonazzi_retina_2024,kang_real-time_2021,wang_multi-sensor_2021}. However, these approaches require significant computing power, which is not practical for battery-operated smart glasses. This constraint restricts their use to applications like gaming and AR-VR, where the eye tracking system remains connected to external devices\cite{akinyelu_convolutional_2020,krishna_raman_2024,frey_gapses_2024,polonelli_h-watch_2021,moosmann_ultra-efficient_2023}. Indeed, despite recent advancements in energy-efficient tinyML, the large amount of generated data requires powerful processors consuming constant power in the range of a few Watts~\cite{akinyelu_convolutional_2020,krishna_raman_2024,das_eog_2024}. To address these challenges, there is a growing interest in developing portable, ultra-low power, and non-invasive eye tracking solutions~\cite{adhanom_eye_2023}. Recent advancements in smart glasses technology have opened up new possibilities for integrating new functionalities into wearable devices~\cite{moosmann_ultra-efficient_2023,frey_gapses_2024}, offering a more convenient, mobile, and less intrusive alternative.


Charge variation (QVar) sensing represents an emerging frontier in biosignal detection, capitalizing on the subtle variations in electric fields generated by biological activity. Unlike traditional EOG, which requires direct skin contact through wet electrodes, this method can operate in a non-contact or minimally invasive manner, enhancing user comfort. By leveraging the corneo-retinal potential—a bioelectric charge created by the retina and cornea during eye movements—this technique enables precise eye tracking with reduced invasiveness. To ensure the efficient processing of these signals within a compact, wearable system, a novel low-power RISC-V processor with an integrated AI accelerator is employed. This processor supports the deployment of tinyML models optimized for real-time performance, achieving high accuracy with minimal latency while maintaining an energy-efficient operation.

This paper introduces ElectraSight, a pair of low-power smart glasses featuring a non-invasive eye tracking system based on QVar sensing. It details the hardware-software co-design, including the proposed tinyML network, and evaluates ElectraSight through comprehensive field experiments.

The key contributions of this work are as follows:
\begin{enumerate*}[label=(\roman*),font=\itshape]
\item A novel usage of integrated QVar sensors for hybrid (contact and contactless) eye tracking. The study fully characterizes and evaluates a novel sensor from STMicroelectronics, featuring a high-impedance differential analog front-end that enables non-invasive, ultra-low power EOG integration on smart glasses. With an average power consumption of \qty{15}{\micro\ampere}, the proposed sensor is capable of detecting subtle electric charge variations, allowing precise eye movement tracking while maintaining a minimal energy footprint, ideal for wearable devices.
\item A low-power miniaturized electronic system designed to completely fit into commercial glasses temples. The designed system is modular and supports up to 6 differential sensing channels, wireless connectivity, and onboard processing based on a novel RISC-V processor. The average measured power consumption is \qty{7.5}{\milli\watt} for data acquisition and \qty{46}{\micro\joule} per movement prediction, enabling continuous real-time operation without sacrificing battery life.  This makes the solution suitable for daily usage.
\item A comprehensive dataset of eye movements, including labeled samples for 10 classes. The dataset serves as a benchmark for evaluating the accuracy of the proposed eye tracking solution and facilitates further research and development in low-power, non-invasive tracking systems.
\item A tinyML model based on a \textit{4-bit} quantized convolutional neural networks optimized for parallel real-time processing of the acquired eye movement data. The model achieves high accuracy while operating within the limited computational resources of the embedded system, making it suitable for deployment on low-power edge devices.
\item Field deployment with real-time onboard operation, demonstrating the practical applicability of the system in real-world conditions. The system successfully operates in live scenarios, showcasing its capability for accurate eye movement classification (up to 92\% accuracy) and real-time feedback (\qty{60}{\milli\second} between movement and detection for 90\% of the movements, \qty{301}{\micro\second} for real-time inference) without the need for external components.
\end{enumerate*}

\section{Related Works}
\label{sec:rel-works}

\begin{table*}[t]
    \caption{Comparison with state-of-the-art (SoA) eye tracking solutions. Results not explicitly reported are marked as Not Specified (N/S). Note *: The latency metric refers to processing latency, which is appropriate for gaze prediction. However, movement classification introduces an additional prediction delay, as the algorithm requires time to observe the signal. COTS is defined as commercial off-the-shelf components.
    }
    \label{tab:soa-comparison}
    \centering
    \renewcommand{\arraystretch}{1.2}
    \begin{tabular}{l c c c c c c c c c c}
        \hline
        \begin{tabular}[c]{@{}c@{}} 
            \textbf{Reference} \\ \textbf{work} \end{tabular} & \begin{tabular}[c]{@{}c@{}}\textbf{Sensing} \\
            \textbf{method}\end{tabular} & \begin{tabular}[c]{@{}c@{}} \textbf{Invasiveness} \\
            \textbf{(L/M/H)}\end{tabular} & \begin{tabular}[c]{@{}c@{}}\textbf{Fully}\\
            \textbf{onboard}\end{tabular} & \begin{tabular}[c]{@{}c@{}}\textbf{Accuracy}\\
           \textbf{ t-($^{\circ}$) / c-(\%)} \end{tabular} & \begin{tabular}[c]{@{}c@{}}\textbf{Field}\\
            \textbf{evaluation}\end{tabular} & \begin{tabular}[c]{@{}c@{}}\textbf{Latency}*\\
            \textbf{(ms)}\end{tabular} & \begin{tabular}[c]{@{}c@{}}\textbf{Average} \\
            \textbf{power [mW]} \end{tabular} & \begin{tabular}[c]{@{}c@{}}\textbf{Price} \end{tabular} & \begin{tabular}[c]{@{}c@{}} \textbf{COTS} \end{tabular} & \begin{tabular}[c]{@{}c@{}}\textbf{Ground} \\ \textbf{Truth} 
        \end{tabular}  \\

        \hline
        PL Neon$^\dagger$    & IR camera  & L & \xmark & t-1.8 & \cmark & N/S (L$^\star$) & N/S (H$^\star$) & \$\$\$\$ &\cmark & N/S\\
        \cite{bonazzi_retina_2024}    & Event camera  & M & \xmark & t-3.24px$^\Delta$ & \cmark & 6 & 5 & \$\$\$\$ &\xmark & \xmark\\
        \cite{das_eog_2024}    & EOG  & H$^\diamond$ & \xmark & c-97.9 & \xmark & N/S & 1446 & \$\$\$ &\cmark & \xmark\\
        \cite{frey_gapses_2024} & 	EOG/EEG  & M & \cmark & c-95 & \cmark & 4000 & 26.5 & \$  &\cmark & \xmark\\
        \cite{li_gazetrak_2024}$^\dagger$     & 	Acoustic  & L & \cmark & t-4.9 & \cmark & 10.3 & 95.4 & \$ &\cmark & \cmark\\
        \cite{shi_eye_2023} & 	Ag NW  & L & \xmark & c-97 & \cmark & N/S& N/S& \$\$\$ &\xmark & \xmark\\
        \cite{guo_wearable_2024}$^\dagger$    & NIR  & L & \xmark & t-1.05 & \cmark & N/S & 670 & \$ &\cmark & \cmark\\
        \rowcolor{customgreen}
        ElectraSight & hEOG  & L & \cmark & c-92/81$^{\star\star}$ & \cmark & 0.30/0-60$^{\star\star\star}$ & 8.85 & \$   &\cmark & \cmark\\ 
        \hline
        \multicolumn{11}{l}{$^\star$Estimated order of magnitude (L, M or H), $^{\star\star}$6 classes / 10 classes, $^{\star\star\star}$Computing time/time between a movement and correct predicition}  \\
        \multicolumn{11}{l}{$^\diamond$Wet electrodes, $^\Delta$Results provided in pixels, 
        $^\dagger$ User-dependent model and/or calibration.}
    \end{tabular}
\end{table*}

Eye tracking technologies have undergone extensive development, with camera-based and EOG-based systems being the most commonly adopted approaches~\cite{akinyelu_convolutional_2020,skaramagkas_review_2023,klaib_eye_2021, das_eog_2024}. This section explores both established and emerging eye tracking technologies, focusing on their advantages, limitations and recent innovations~\cite{skaramagkas_review_2023}, particularly in low-power solutions~\cite{bonazzi_retina_2024}.

\subsection{Eye Tracking}
\subsubsection{Camera based}
Eye tracking systems have traditionally been dominated by camera-based systems~\cite{bonazzi_retina_2024,akinyelu_convolutional_2020,skaramagkas_review_2023,klaib_eye_2021}, which offer high accuracy but at the cost of significant power consumption and high computational demand. These systems rely on infrared (IR) cameras to monitor eye movements and estimate gaze direction~\cite{skaramagkas_review_2023}. Commercial products like the Tobii Pro Fusion and Pupil Labs (PL) Neon are widely used for research and industry applications. The Tobii Pro Fusion, for instance, can be attached to a display to assess gaze behavior. At the same time, wearable solutions like the PL Neon glasses feature dual IR cameras capturing eye data at \qty{200}{\hertz}. The PL Neon achieves a per-subject median accuracy of \ang{1.8} across a \ang{60} \(\times\) \ang{30} field of view when calibrated, improving to \ang{1.3} after fine-tuning. However, these systems are tethered to external devices, such as smartphones or computers, due to their energy and processing requirements, which also contribute to high cost and limited portability~\cite{skaramagkas_review_2023,shi_eye_2023}.
Video-based systems also raise privacy concerns, as they capture sensitive visual data that may unintentionally expose personal information~\cite{kroger_what_2020}. The need for external devices for processing or storage can further increase the risk of unauthorized access or data breaches, making users hesitant to adopt these technologies.
The work presented in~\cite{bonazzi_retina_2024} utilizes compact, low-power event-based cameras capable of detecting image movements to perform eye tracking. This approach is promising, achieving pixel-level accuracy (3.24 pixels) and demonstrating low theoretical power consumption (\qty{5}{\milli\watt}). However, the prototype relies on components that are not yet commercially available or prohibitively expensive, as this technology is still in its early stages of development and evolution. The computing is also not integrated onboard the frames.

\subsubsection{EOG based}
In contrast, \textit{EOG-based systems} utilize electrodes placed around the eyes to detect electrical potentials generated by eye movements. Although generally less power-hungry than camera-based solutions, EOG systems face challenges in accuracy and ease of use, with baseline drift, being a significant issue~\cite{barbara_comparison_2020}.  This drift interferes with the acquired data, causing signal shifts even when the eyes remain stationary. Accurate predictions above 90\% are typically achieved only in controlled environments with wet electrodes and minimal movement~\cite{lopez_comparison_2023}. As a result, robust eye tracking remains difficult, and non-invasive EOG systems primarily focus on eye movement detection\cite{lopez_comparison_2023}. These systems usually evaluate movement classes such as \textit{blink}, \textit{left}, or \textit{up}, in contrast to camera-based approaches, which measure eye gaze as angles on the x and y axes~\cite{bonazzi_retina_2024}.
In~\cite{das_eog_2024}, a dataset acquired using five invasive wet electrodes on six classes, namely \textit{no movements, down, up, left, right, blink} is used. A Machine Learning (ML) algorithm runs on an FPGA, reaching an accuracy of 95.56\%. However, despite running on an embedded platform, the system consumes \qty{1446}{\milli\watt} and is not designed for integration into glasses frames. While the system demonstrates a high accuracy, it remains expensive, invasive, and lacks on-field validation.
Aiming for a fully onboard solution,~\cite{frey_gapses_2024} presents a smart glasses system that combines EOG and electroencephalography (EEG) with dry electrodes. Aiming for a fully onboard solution,~\cite{frey_gapses_2024} presents a smart glasses system that combines EOG and electroencephalography (EEG) with dry electrodes to classify 11 movement types, achieving 95\% accuracy with a power consumption of \qty{26.5}{\milli\watt}. However, this system is also relatively invasive, requiring user-specific model training. The glasses utilize semi-rigid dry electrodes, including brush pins, which can cause irritation and discomfort with prolonged use. Additionally, the authors do not evaluate system latency, relying on four second acquisition windows, which limits the system's ability to detect quick and successive movements.

\subsubsection{Others}

Other approaches have also been explored, each offering distinct advantages and facing specific limitations.
In~\cite{shi_eye_2023}, a system that uses electrostatic interfaces integrated into glasses is proposed, achieving 97\% classification accuracy on nine categories of eye movements. This system leverages the benefits of non-contact sensing to avoid the need for electrodes to touch the skin, thus making it more comfortable and suitable for daily use. However, the latency and power consumption are not quantified, and the system relies on custom-made lens electrodes, which complicates the industrialization process. The work in~\cite{guo_wearable_2024} utilizes ultrasound acoustic measurements with microphones and speakers, offering a high accuracy of \ang{4.9} while being non-invasive and fully integrated. The entire pipeline is evaluated, with a reported power consumption of \qty{95.4}{\milli\watt} for live gaze estimation. A potential limitation of this system is its susceptibility to external acoustic noise, such as wind. 
An affordable system based on near-infrared (NIR) LED and phototransistors is presented in~\cite{das_eog_2024}. The work reaches an accuracy of \ang{1.05} with a power of \qty{670}{\milli\watt} for sensing only. However, the processing is not done in real-time or onboard.

All the works presented above are summarized with key parameters in \Cref{tab:soa-comparison}.
In summary, integrating eye tracking into non-stigmatizing smart glasses remains challenging due to the absence of solutions that combine high accuracy, low power consumption, non-invasiveness, affordability, cross-user compatibility, low latency, the ability to detect multiple movements per second, and reliance on commercially available components. This paper aims to address these challenges by presenting a solution that bridges this gap.

\subsection{QVar Applications}

QVar is an electrical potential sensing technique that measures quasi-electrostatic potential changes using a high input impedance differential amplifier. More details are provided in \Cref{sec:qvar-sensor}. STMicroelectronics recently released a series of sensors with ultra-low power QVar capabilities and a small footprint. Its applications are still being explored. Di Fiore et al. \cite{di_fiore_innovative_2023} used the QVar sensor to estimate the airspeed of drones by measuring the electrostatic charge that builds up due to friction with the surrounding air. In gesture recognition, Reinschmidt et al. \cite{reinschmidt_realtime_2022} demonstrated that combining QVar and IMU data can improve classification accuracy by more than 10\% compared to IMU-only systems. In~\cite{dheman_towards_2023}, QVar sensors are used to detect electrocardiogram (ECG) signals and heartbeats, creating a low-power alternative for heart rate monitoring devices that consumes only \qty{87.3}{\micro\watt}. Further, they compared QVar to traditional electromyography (EMG) sensors in bladder monitoring, finding a high correlation between the two, suggesting that QVar could be a viable low-power replacement in biomedical applications. Schulthess et al. \cite{schulthess_tinybird-ml_2023} applied QVar in animal-borne sensor nodes to monitor birds' vital signs, such as heartbeats and muscle contractions.

While QVar finds application in various domains, its use in eye tracking remains largely unexplored. Manoni et al. \cite{manoni_long-term_2022} investigated QVar for biopotential measurement, comparing its performance to a laboratory gold-standard device for ECG, EEG, and EOG signals. However, the EOG study is limited to rapid eye movement detection and did not explore full eye tracking. This paper aims to address this gap by evaluating the QVar sensor for low-power eye tracking, focusing on both contact and non-contact sensing approaches.

\section{System Architecture}
\label{sec:electronics}
This section presents the system architecture including the concept of capacitive and electrostatic charge variation sensing, the description of QVar sensors, the impedance characterization and model of the system as well as the hardware setup used for this work.

\subsection{Capacitive and Electrostatic Charge Variation Sensing}

Capacitive sensing is an emerging field in biomedical and environmental sensing that measures biological and electrical signals by detecting changes in charge using highly sensitive devices. This technique exploits the ubiquitous presence of electric fields, measuring their effects on charges~\cite{forouhi_cmos_2019}. 



\begin{figure}
    \centering
    \includegraphics[width=0.9\linewidth]{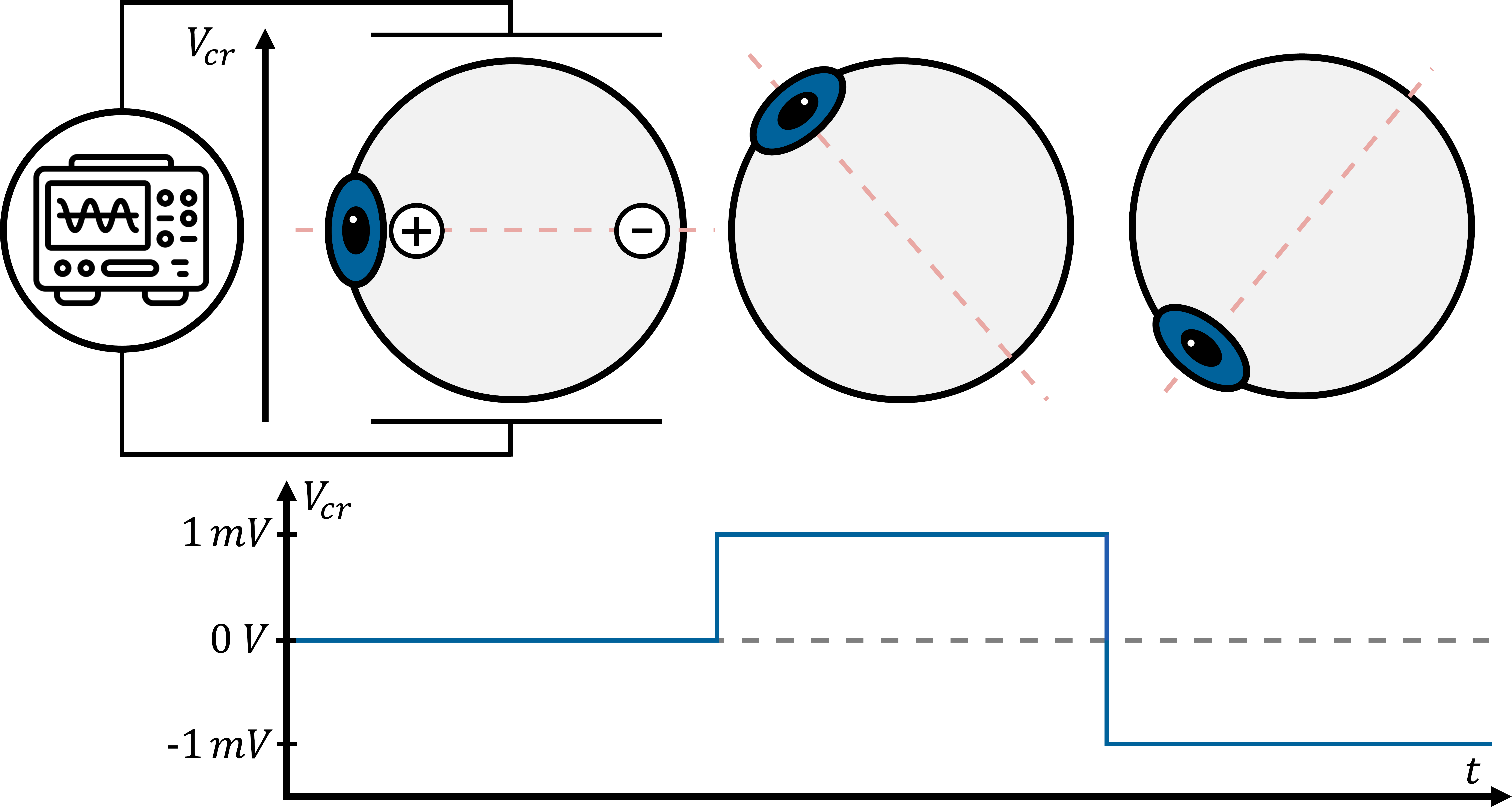}
    \caption{Corneo-retinal potential, typically spanning between 250 and \qty{1000}{\micro\volt} with a bandwidth of $\sim$\qty{30}{\hertz} 
~\cite{brown_iscev_2006}.}
    \label{fig:cpr}
\end{figure}

This work presents an eye-movement detection technique based on QVar sensing. A combination of galvanic (contact-based) and capacitive (contactless) sensing is employed to integrate the system into the frame of standard eyeglasses. The biological signal of interest is the corneo-retinal potential~\cite{brown_iscev_2006}, a bioelectric charge generated by the retinal cells during normal visual function. The typical corneo-retinal potential spans between 250 and \qty{1000}{\micro\volt} with a bandwidth between 0.5 and \qty{30}{\hertz}~\cite{brown_iscev_2006}. The cornea, located at the front of the eye, creates a measurable electrical potential difference that can be detected during eye movements. This phenomenon is illustrated in \Cref{fig:cpr}. 

\subsection{Onboard Sensors} 
\label{sec:qvar-sensor}

As discussed in \Cref{sec:rel-works}, QVar sensing detects changes in quasi-electrostatic potential using a high-impedance differential amplifier. STMicroelectronics offers an integrated circuit family that combines amplification, digitization, and signal transmission into a single solution. A QVar sensor has two inputs with differential electrode connections, which are linked to a high-impedance analog front-end (AFE) operating in the giga-ohm range. The AFE provides biasing and amplification for the electrodes, after which the signal is digitized by an integrated analog-to-digital converter (ADC). Finally, the processed signal is read by a host device, typically a microcontroller unit (MCU).

This work uses two sensors from STMicroelectronics: the LSM6DSV16X, which integrates a 3-axis accelerometer and gyroscope, and the ST1VAFE3BX, an ultra-low-power accelerometer. Both sensors feature a charge variation sensing channel and support in-sensor processing for tasks such as measurement filtering, step detection, and state machine execution. Their extensive features and configurability make them ideal for highly integrated applications, especially where sensor fusion can provide a benefit. For instance, movement detected by the accelerometer can help filter out movement noise from the QVar channel.

The QVar channel in the LSM6DSV16X IMU features a configurable input impedance ranging from \qty{235}{\mega\ohm} to \qty{2.4}{\giga\ohm} and operates with an input dynamic range of \qty{\pm 460}{\milli\volt}. The channel has a sensitivity of \qty{78}{LSB\per\milli\volt} and a fixed sampling frequency of \qty{240}{\hertz}, with an input noise of \qty{54}{\micro\volt}. The LSM6DSV16X specifications are summarized in \Cref{tab:qvar_characteristics}. This sensor is therefore suitable for detecting subtle electrical changes, such as those associated with touch proximity sensing or biosignals. Additionally, the QVar channel consumes approximately \qty{15}{\micro\ampere} at \qty{1.8}{\volt}. However, it is not possible to isolate and use the QVar channel independently while deactivating the IMU, bringing the total sensor power consumption to \qty{650}{\micro\ampere}.
The ST1VAFE3BX shares similar features but introduces several key improvements: programmable gain, a higher sampling frequency (up to \qty{3200}{\hertz}), and a lower total current consumption (\qty{48}{\micro\ampere}).

A multi-channel device for acquisition is designed to enhance eye tracking performance and prioritize the processing of critical information. This device centers around the LSM6DSV16X sensor, which is used for dataset collection and channel position evaluation. The newer and optimized ST1VAFE3BX, released later, is used in the final prototype depicted in \Cref{fig:inference-system-description}.

\begin{table}[t]
\centering
\caption{Electrical characteristics of the LSM6DSV16X QVar channel.}
\begin{tabular}{ll}
    \hline
    \textbf{Characteristic} & \textbf{Value} \\
    \hline
    Power consumption & \qty{15}{\micro\ampere} \\
    Input impedance (R1 - R2) & \qty{235}{\mega\ohm} to \qty{2.4}{\giga\ohm} \\
    Input range & \qty{\pm 460}{\milli\volt} \\
    Sensitivity & \qty{78}{LSB\per\milli\volt} \\
    Sampling frequency & \qty{240}{\hertz} \\
    Noise level & \qty{54}{\micro\volt} \\
    \hline
    \\
\end{tabular}
\label{tab:qvar_characteristics}
\end{table}

\begin{figure}
    \centering
    \includegraphics[width=1\linewidth]{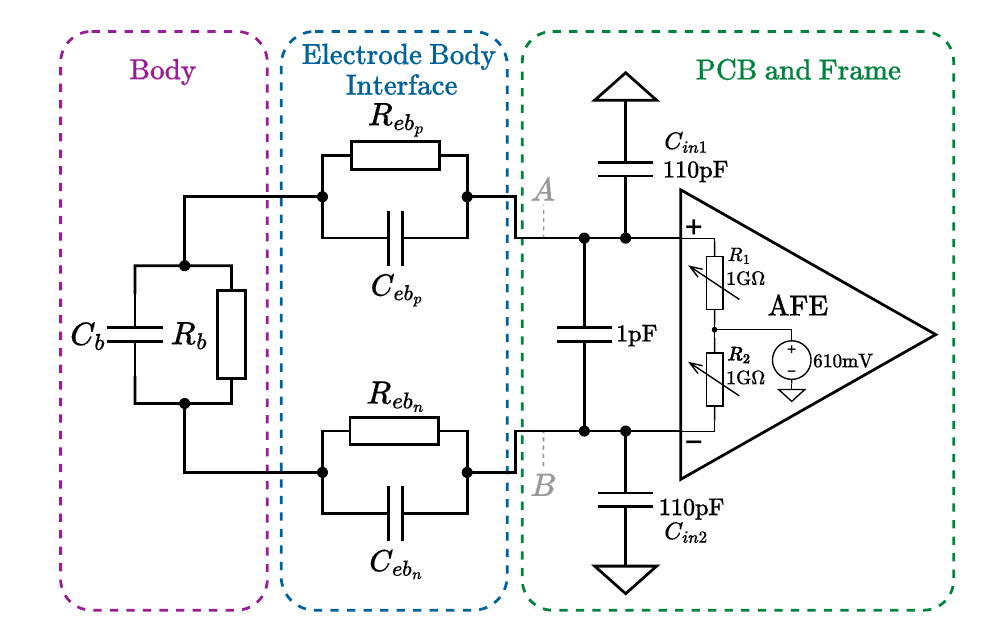}
    \caption{Equivalent model of sensor - body interface}
    \label{fig:impedancemodel}
\end{figure}
\subsection{System Impedance Characterization}
\label{sec:impedance}
To verify the feasibility of non-invasive contactless eye tracking, the impedance of the entire sensor-frame-body interface is characterized. This analysis enables essential QVar characterization to accurately measure the corneo-retinal potential via contactless electrodes, which results in a high output skin impedance for each specific differential channel pair. To prevent further attenuation of the faint EOG signal (below \qty{1}{\milli\volt}), the QVar input impedance must exceed that of the sensor-frame-body interface by several orders of magnitude.

\Cref{fig:impedancemodel} illustrates a simplified equivalent circuit diagram for the sensor-frame-body system. While STMicroelectronics does not provide an internal circuit diagram for the LSM6DSV16X sensor, its behavior and characterization suggest it can be modeled as an instrumentation amplifier with giga-ohm programmable input impedance (R1 and R2). The input to each QVar channel begins at the printed circuit boards (PCB) and frame section, where a \qty{110}{\pico\farad} stabilization capacitor is used, acting as a low-pass filter when combined with the resistive electrode-body interface. An input protection circuit is also included (not shown). The total parasitic capacitance between PCB traces and cables on the frame is approximately \qty{1}{\pico\farad}.

The body impedance of all combinations of interfaces is measured using a \textit{Keysight E9480AL} LCR meter, combined with \textit{16089C} Kelvin IC clip leads attached to points A and B of the equivalent schematic shown in \Cref{fig:impedancemodel}. All combinations of impedances are measured in full equivalent schematic of the AFE of the QVar channel with the most relevant parasitic capacitances, interface, and body impedances shown in \Cref{tab:impedance}.

The frame-body attenuation derived from \Cref{fig:impedancemodel} and \Cref{tab:impedance} confirms that the signal attenuation is negligible for contact electrodes in the 0.5 and \qty{30}{\hertz} bandwidth. With a body resistance in the range of $\sim$\qty{10}{\kilo\ohm} ($R_b$) and the electrode series \qty{50}{\kilo\ohm} - \qty{200}{\kilo\ohm} $R_{eb}$, the attenuation ratio is around 8\textperthousand. On the other hand, the series capacitance for contactless sensing plays a role in signal attenuation. Indeed, $C_{eb}$ adds a series impedance of \qty{1}{\giga\ohm}, which, combined with the maximum input impedance derived from \Cref{tab:qvar_characteristics}, gives an attenuation factor of 0.7. In the worst case, selecting the minimum input impedance of \qty{235}{\mega\ohm}, the attenuation factor would degrade to 0.19. Therefore, the typical corneo-retinal potential is perceived scaled down between 175 and \qty{700}{\micro\volt}~\cite{brown_iscev_2006}. With a noise level of \qty{54}{\micro\volt} and a sensitivity down to \qty{13}{\micro\volt}, the proposed electronics is capable of measuring EOG signal with full contactless electrodes mounted on glasses rims; therefore, demonstrating the feasibility of non-invasive real-time eye tracking monitoring.    
\begin{table}[t]
\centering
\caption{Estimated parameters for each electrode pair from \Cref{fig:electrode-position}.  DRT--DRN: \textit{Diagonal Right Temple}, \textit{Diagonal Right Nose}. DLT--DLN:\textit{ Diagonal Left Temple}, \textit{Diagonal Left Nose}. CL--CR (contactless): \textit{Center Left}, \textit{Center Right}. HL--HR (contactless): \textit{Horizontal Left}, \textit{Horizontal Right}. VT--VB (contactless): \textit{Vertical Top}, \textit{Vertical Bottom}.}
\begin{tabular}{lcccccc}
\hline
\textbf{Pair} & \ensuremath{\mathbf{R_{eb_p}}} & \ensuremath{\mathbf{C_{eb_p}}} & \ensuremath{\mathbf{R_b}} & \ensuremath{\mathbf{C_b}} & \ensuremath{\mathbf{R_{eb_n}}} & \ensuremath{\mathbf{C_{eb_n}}} \\

\hline
DRT--DRN & 50\,k$\Omega$ & 20\,nF & 16\,k$\Omega$ & 100\,nF & 200\,k$\Omega$ & 20\,nF \\
DLT--DLN & 50\,k$\Omega$ & 25\,nF & 16\,k$\Omega$ & 100\,nF & 150\,k$\Omega$ & 25\,nF \\
CL--CR   & - & 3\,pF & 30\,k$\Omega$ & 75\,nF & -& 3\,pF \\
HL--HR   & - & 3\,pF & 8\,k$\Omega$ & 20\,nF & - & 3\,pF \\
VT--VB   & - & 20\,pF & 8\,k$\Omega$ & 20\,nF & - & 4\,pF \\ \hline
\end{tabular}
\label{tab:impedance}
\end{table}

\subsection{Hardware Architecture}
A major contribution of this paper is the investigation of contact and contactless differential channels for detecting eye movements directly onboard the glasses, achieving high accuracy and low latency. To this end, an ultra-low-power, battery-powered electronic system is developed. It acquires data from six QVar channels and is able to transmit it via Bluetooth Low Energy (BLE). As no commercially available solutions meet these requirements, custom PCBs are designed based on an open-source project named VitalCore\footnote{\url{github.com/ETH-PBL/VitalCore}}, a technical choice that supports repeatability and expansion for future works. As depicted in \Cref{fig:electronics-block-diagram}, the system consists of three main blocks, VitalCore~\cite{ronco_tinyssimoradar_2024}, tinyML VitalPack, and QVar VitalPack. 

\Cref{fig:electronics-3d} represents a 3D view of the three main boards and their main components. As it is, the hardware design does not reach optimal compactness and minimal dimensions achievable with an engineered single PCB design. This intentional technical decision finds the optimal trade-off between electronic dimension and system flexibility, a key element in research. For example, sub-parts of the system can be modified without requiring a complete electronic redesign. The three boards described in more detailed in the following section.

\subsubsection{VitalCore}
The heart of the system is the VitalCore~\cite{ronco_tinyssimoradar_2024}, an open-source platform specifically designed for low-power wearable projects. This highly integrated, miniaturized embedded system provides all the necessary components for a battery-supplied and wireless sensor node. Its compact dimensions of \(\qty{17.6}{\milli\meter} \times \qty{12.6}{\milli\meter}\) make it an ideal foundation for space-constrained designs, a crucial factor for the target application.
The platform is built around the NRF5340  system on a chip (SoC), featuring a dual-core Cortex-M33 processor with a maximum clock speed of \qty{128}{\mega\hertz}, \qty{1}{\mega\byte} of Flash memory, \qty{512}{\kilo\byte} of RAM, and Bluetooth 5.2, along with an integrated chip antenna. On the bottom PCB side, the VitalCore is equipped with a \qty{0.4}{\milli\meter} pitch, 50-position connector, facilitating the connection of application-specific 'VitalPack' expansion boards. This connector provides access to power inputs and outputs, the SoC’s programming interface, USB port, and 28~GPIOs pins that support a variety of interfaces. 

\subsubsection{tinyML VitalPack}

To ensure low-latency execution—within the millisecond range—and high-accuracy tinyML model performance, a custom VitalPack PCB is designed to handle heavy computational loads while maintaining a compact form factor (\(19.5 \times 16.5 \times \qty{0.8}{\milli\meter}\)). The main component is the GAP9 from Greenwaves, a low-power microcontroller designed for edge AI applications where energy efficiency is crucial. In the presented custom hardware, it works as a coprocessor. The communication between the NRF5340 and the GAP9 is handled via a snap connector.

The GAP9 SoC is based on open hardware, namely the PULP platform and a RISC-V instruction set architecture (ISA) and features a multi-core architecture that balances performance and power consumption, embedding 10 RISC-V-based cores organized into two main power and frequency domains. The first domain, known as the fabric controller (FC), has a single core running at up to \qty{400}{\mega\hertz} and comes with \qty{1.5}{\mega\byte} of SRAM (L2 memory). The FC manages communications with peripherals and coordinates memory operations.
The second domain, referred to as the Cluster (CL), comprises nine RISC-V cores, each operating at up to \qty{400}{\mega\hertz}. The CL is designed for tasks that require parallel processing, with \qty{128}{\kilo\byte} of L1 zero-latency memory shared by the CL cores. A key feature of the GAP9 is its neural engine, which speeds up operations like convolutions, batch normalization, and ReLU activations. This engine is designed to handle quantized deep learning models, making the processor well-suited for complex ML tasks. In this paper, the clock speed is set at \qty{370}{\mega\hertz} for the CL and the FC.

The GAP9 is highly efficient, it can perform up to 2-Tera operations per second (TOPS) while consuming an average of \qty{170}{\milli\watt}, making it ideal for battery-powered devices. When only the CL is operative, the power consumption decreases to \qty{70}{\milli\watt}. Additionally, the GAP9 features smart power management that dynamically adjusts power usage based on workload, extending battery life and reducing energy consumption. These capabilities make the GAP9 particularly well-suited for the scope of this paper.

\subsubsection{QVar VitalPack}
\begin{figure}[t]
\centering
\begin{subfigure}{0.475\textwidth}
    \centering
    \includegraphics[width=1\linewidth]{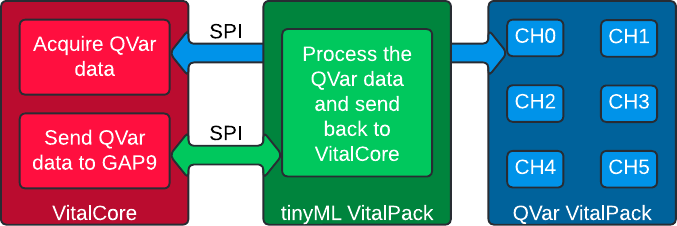}
    \caption{Block diagram of the electronics system. The nRF53 acquires data from the QVar sensors and forwards it to the GAP9. The GAP9 then sends the result of the inference back to the NRF53.}
    \label{fig:electronics-block-diagram}
\end{subfigure}
\begin{subfigure}{0.475\textwidth}
    \centering
    \includegraphics[width=1\linewidth]{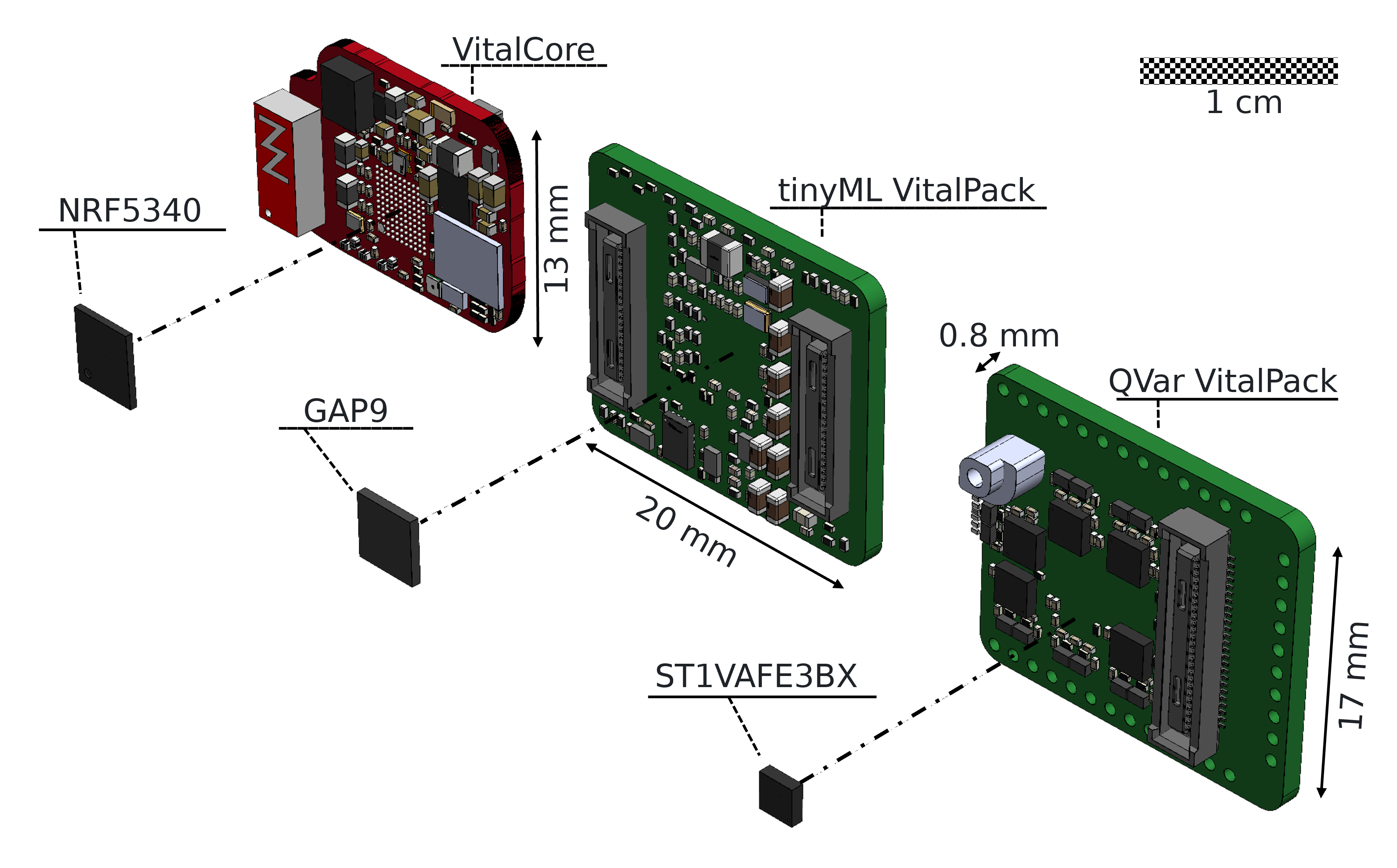}
    \caption{Exploded view of the three main electronic blocks: namely VitalCore, tinyML VitalPack, and QVar VitalPack. PCB dimensions and the three main components are highlighted.}
    \label{fig:electronics-3d}
\end{subfigure}
\caption{Logical block diagram (a) and 3D exploded view (b) of the electronic used to enable plug-and-play non-invasive eye tracking.}
    \label{fig:electronics}
\end{figure}

A VitalPack featuring six QVar channels is developed, providing a flexible system that allows for the testing and implementation of various electrode position combinations. Its compactness of \(\qty{19.5}{\milli\meter} \times \qty{16.5}{\milli\meter}\) makes it relevant for the final design.
The QVar VitalPack is based on six \textit{ST1VAFE3BX} sensors, as described in \Cref{sec:qvar-sensor}. 
The QVar VitalPack is connected to the rest of the system using the same connector as on the aforementioned boards. It is plugged into the bottom side connector of the GAP VitalPack but only communicates with the VitalCore,  utilizing the GAP9 connector as a pass-through.
The whole electronics system and its integration are described in \Cref{fig:electronics-block-diagram} and \Cref{fig:electronics-3d}.

\subsubsection{System Interaction}
The NRF53 on the VitalCore is connected to the six QVar sensors using SPI. The acquired data is read by the NRF53 using direct memory access (DMA) and the FIFO to optimize power consumption.
The NRF53 then splits the collected QVar data points in predefined windows with fixed time lengths, which are subsequently forwarded to the GAP9 co-processor over SPI.

\section{Smart Glasses - System Setup}
\label{sec:system}
\begin{figure}
    \centering
    \includegraphics[width=1\linewidth]{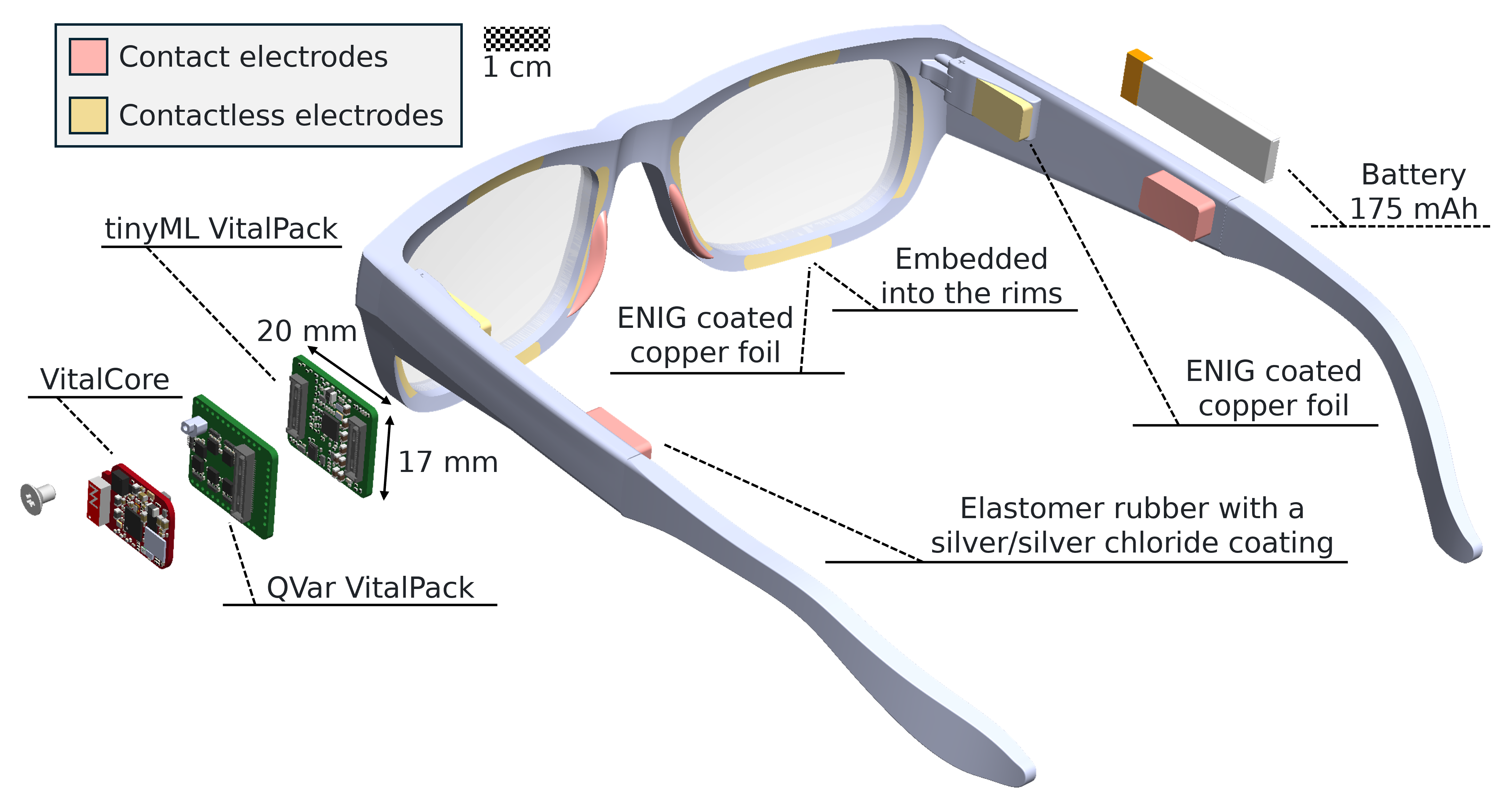}
    \caption{3D model of the final prototype. The electronics are on the left temple, and the \qty{175}{\milli\ampere\hour} battery is on the right one. The electronics and electrodes are connected through wires inside the glasses frame.}
    \label{fig:inference-system-description}
\end{figure}

ElectraSight incorporates the electronics described in \Cref{sec:electronics}, seamlessly embedded within the temples of the eyeglasses. The system utilizes five differential channels and is powered by a \qty{175}{mAh} lithium battery.

Two distinct prototypes are developed. The first prototype, designed primarily for data acquisition, is more versatile and incorporates an eye tracker and is based on the LSM6DSV16X. The second prototype, a more compact and non-stigmatizing final version, utilizes the ST1VAFE3BX sensor for real-time inference. The latter is shown in \Cref{fig:inference-system-description}. 
The system is fully focused on pervasiveness for daily usage and scalability, providing a truly \textit{plug-and-play} solution that does not need any initial calibration, electrode placing, or invasive installation procedure. 

Thorough investigation and field testing led to the selection of two types of electrodes: one for contact sensing and the other for contactless sensing. 
\textit{Softpulse} electrodes from \textit{Dätwyler} for contact sensing are made from a conductive elastomer rubber with a silver/silver chloride coating, offering biocompatibility for safe use on the skin. Their flexibility and low skin impedance further enhance the quality of contact with the skin, ensuring reliable signal acquisition~\cite{van_der_heijden_multi-channel_2024}. 
For the contactless electrodes, sheets of copper coated with an immersion gold (ENIG) layer are used due to their versatility, ease of use, and ability to provide a strong coupling with the human skin. Since these electrodes normally do not operate in direct contact with the skin, they can be directly integrated into the glasses frame. 

The system employs five channels, which requires a total of ten electrodes, as described in \Cref{fig:electrode-position,fig:inference-system-description}.
Two pairs of electrodes serve as contact electrodes, utilizing the standard contact points of eyeglasses at the nose pads and the front of the temples. The first pair, referred to as \textit{diagonal left}, connects the left temple electrode to the left nose pad. The second pair, \textit{diagonal right}, connects the right temple electrode to the right nose pad. These two channels are nearly orthonormal and enable distinct signals to be observed in both channels when the wearer performs different eye movements. 
Three pairs of electrodes are contactless, each consisting of rectangular copper sheets measuring \(\qty{2.5}{\milli\meter} \times \qty{0.7}{\milli\meter}\). These electrodes are positioned around the left eye to form \textit{vertical} and \textit{horizontal} \textit{channels}. The \textit{vertical channel} electrodes are horizontally aligned with the center of the eye without impacting the user's field of view. One electrode is positioned between the upper eyelid and the eyebrow, while the other is located between the lower eyelid and the cheekbone. The \textit{horizontal channel} electrodes are aligned with the eye's center and positioned near its horizontal extremities, ensuring they do not obstruct the wearer's field of view. The \textit{center channel} electrodes are positioned on the both temples, close to the rims.

The aforementioned electrode placement is a result of field empirical experiments, finding the best trade-off between the system's invasiveness and performance. The channel selection is the result of the study performed in \Cref{sec:exp-setup}.
As expected from the input impedance characterization in \Cref{sec:impedance}, the best results are achieved using the highest input impedance of \(\qty{2.4}{\giga\ohm}\). 
\section{Experimental Setup and Dataset Acquisition}
\label{sec:exp-setup}
%

To ensure realistic system evaluation and support the training of a lightweight tinyML model, a dataset with ground truth signals and raw sensor data from five QVar channels is recorded. The data is collected from 20 subjects of varying ages and genders performing nine predefined eye movements, such as looking left or blinking. Informed consent is obtained from all participants prior to data collection.

\subsection{Protocol}
The prototype depicted in \Cref{fig:inference-system-description} is installed on the subject, who is instructed to sit in front of a computer screen, at a distance of approximately \qty{60}{\centi\meter}. Parameters such as the subject's distance to the screen, ambient room illumination, and any relevant observations or comments from the subject are documented.
The subject is instructed to minimize body movement to prevent artifacts in the measured signals and to avoid blinking, except when directed otherwise. To facilitate this, breaks are scheduled throughout the acquisition to allow for relaxation of the eyes. 

\begin{figure}
    \centering
    \includegraphics[width=.9\linewidth]{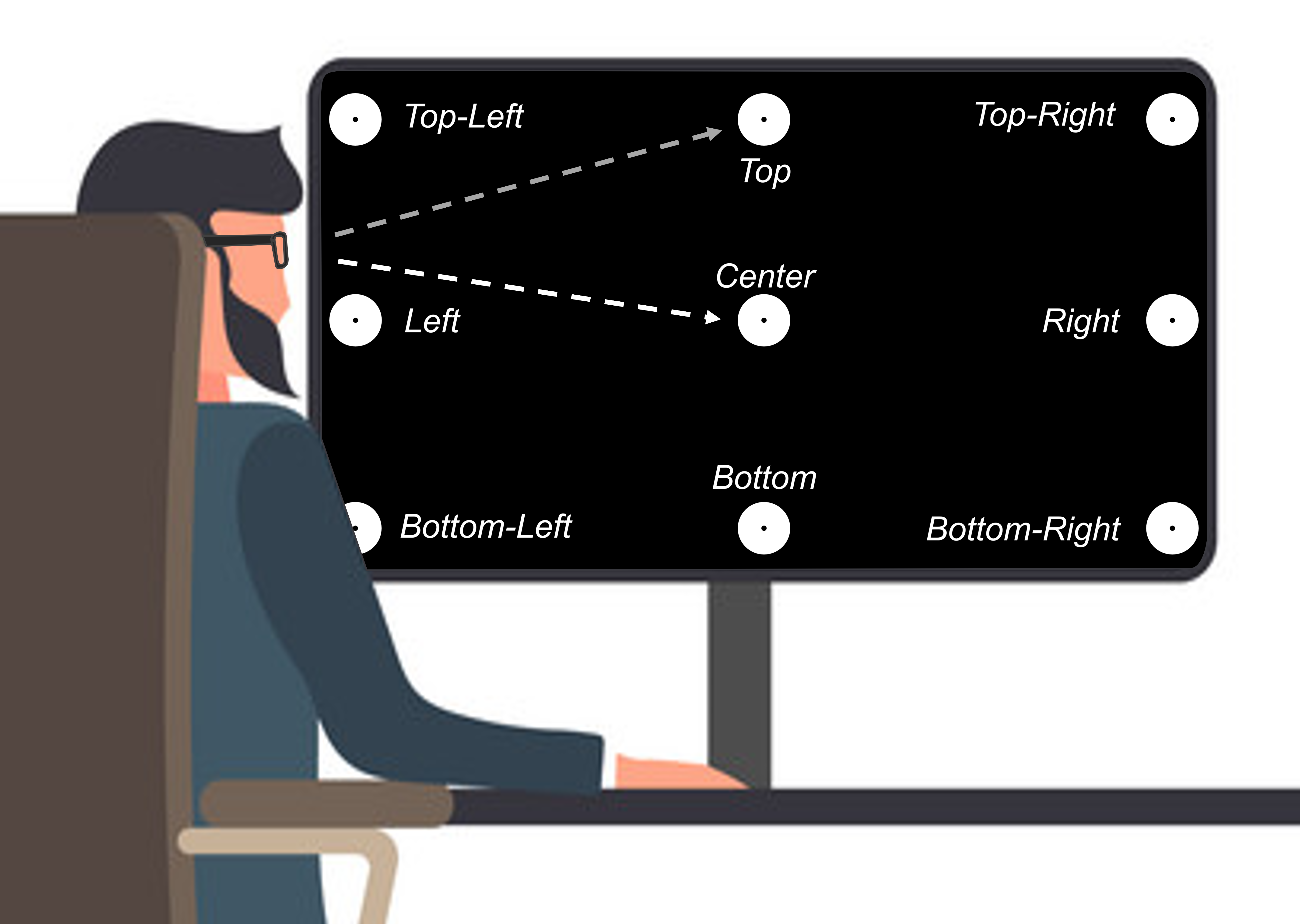}
    \caption{Dataset acquisition setup}
    \label{fig:dataset-positions}
\end{figure}

Each session consists of five recordings. A recording is a sequence of instructions displayed on the computer screen. First, the user is asked to perform four blinks with an interval of one second and to stare for five seconds in the middle of the screen. This serves for synchronization during data analysis. Then, a dot moves repeatedly between eight positions and returns to the center of the screen at one-second intervals. To capture blink events, the word "Blink" appears at the center of the screen. Each position is displayed twice in a random pattern to prevent subject anticipation. Those positions are to the top, bottom, left, right, and all four corners of the screen, as shown in \Cref{fig:dataset-positions}. Since five recordings are carried out per subject, each subject repeats each movement ten times. The time at which the user is prompted to perform a movement is recorded as \textit{logger label}.

\subsection{Ground Truth Eye-Tracker}
\begin{figure}[t]
\centering
\begin{subfigure}[t]{0.46\linewidth}
    \centering
    \includegraphics[width=\linewidth]{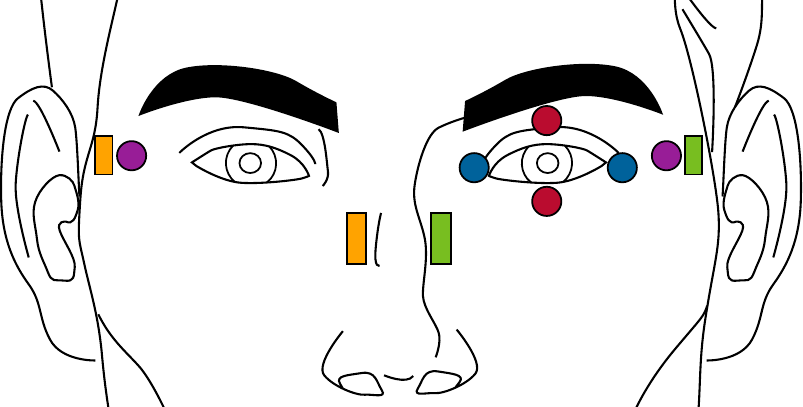}
    \caption{Electrode placement.}
    \label{fig:electrode-position}
\end{subfigure}
\begin{subfigure}[t]{0.46\linewidth}
    \centering
    \includegraphics[width=\linewidth]{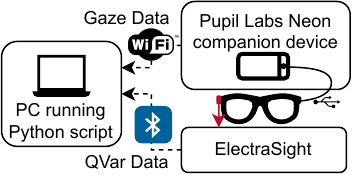}
    \caption{Experimental setup.}
    \label{fig:exp-setup-description}
\end{subfigure}
\hfill
\caption{(a) Experimental setup description of the five differential channels in the glasses setup: \textit{diagonal left} (orange), \textit{diagonal right} (green), \textit{horizontal} (blue), \textit{vertical} (red), and \textit{center} (purple); with contact (rectangles) and contactless (dots) electrodes, and (b) description of the acquisition system setup.}
\label{fig:exp-setup-and-electrodes}
\end{figure}

The Pupil Labs Neon glasses are used as a reference system. It features two IR cameras that record the eye at \qty{200}{\hertz}. The gaze estimation is then done in real-time on a smartphone, called Companion Device at a frequency of \qty{120}{\hertz}. They are considered the ground truth since they have an absolute accuracy better than \ang{1.8} in both axes.
The Companion Device streams the gaze data over the local WiFi network to the acquisition computer as shown in \Cref{fig:exp-setup-description}. 

\section{Data}
\label{sec:data}
%
The eye-tracker and the QVar data for those recordings are gathered in a dataset. This chapter provides details on this data, and how it is synchronized and processed.

The provided gaze data consists of the azimuth and elevation angles as described in \Cref{fig:labeled-qvar-gaze}. The azimuth describes the horizontal angle while the elevation describes the vertical angle. The coordinates are \((0,0)\) when the subject looks straight ahead and become positive when looking higher or to the right. The QVar data is captured at \qty{240}{\hertz} across five channels. An example of raw and filtered QVar data is shown in \Cref{fig:labeled-qvar-gaze}.



\subsection{Movement Labeling}
To achieve accurate labeling, eliminate incorrect movements, and precisely identify their timing, logger labels are combined with eye-tracker gaze data. As mentioned in \Cref{sec:exp-setup}, the logger data includes ten movement classes: \begin{enumerate*}[label=(\roman*),,font=\itshape]
    \item \textit{up}~(U),
    \item \textit{down}~(D),
    \item \textit{left}~(L),
    \item \textit{right}~(R),
    \item \textit{down-left}~(DL),
    \item \textit{up-left}~(UL),
    \item \textit{up-right}~(UR),
    \item \textit{down-right}~(DR),
    \item \textit{straight}~(S),
    \item and \textit{blink}~(B).
\end{enumerate*}

These first eight movements correspond to the subject looking from the center of the screen to specific points illustrated in \Cref{fig:dataset-positions}. However, return movements—where the subject looks back from a point to the center—are also captured. Since the classification focuses on directional movements, an eye movement from the center to the right is considered similar to one from the left to the center. These movements are used to double the number of samples for the movement classes. \textit{Basic movements} are all of the movements without taking the corners into account, while the \textit{full movements} are taking all of the movements into account.

\begin{figure}[t]
    \centering
    \includegraphics[width=1\linewidth]{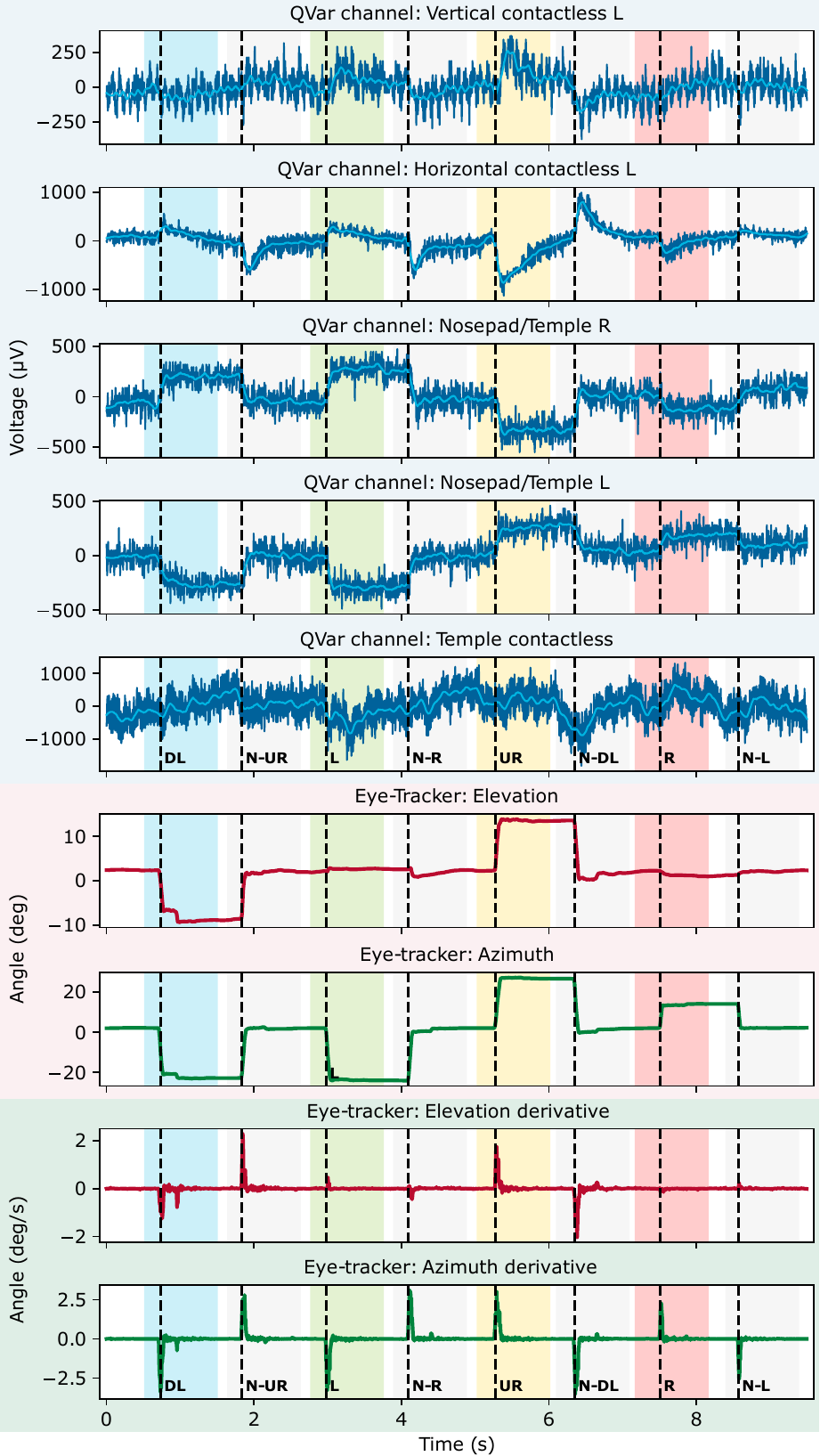}
    \caption{Example raw (blue) and Savitzky-Golay-filtered (cyan) QVar data for all the channels. Gaze (elevation and azimuth) data from the ground truth eye-tracker and its derivative, used for precise labeling. Labels are represented using the logger-based approach (color boxes) as well as the eye tracking-based algorithm (dotted line with the corresponding). The text corresponds to the movement labels, with \textit{N} standing for \textit{Negative} movements.}   
    \label{fig:labeled-qvar-gaze}
\end{figure}

The initial labeling step involves calculating the derivative of the gaze data for elevation and azimuth. Using the logger data, segments of one second following predefined movements are created from the derivative signal. These segments are then used to compute the mean and standard deviation of the gaze derivative, applying standardization as shown in \Cref{eq:standardization}, where \( X \) represents the raw gaze data, \( \mu \) the mean, \( \sigma \) the standard deviation and \( Z \) the standardized gaze data.
\begin{equation}
Z = \frac{X - \mu}{\sigma}
\label{eq:standardization}
\end{equation}
%
Next, the remaining movement segments are evaluated using a threshold-based approach on the gaze derivative ($Elev^{'}$ and $Az^{'}$). The thresholds are defined in \Cref{eq:thresholds}, which are selected based on manual inspection of the gaze derivative data. 
\begin{equation}
\begin{split}
\left\{\begin{matrix}
 Th_{up} & : & Elev^{'} & > & 3 \\
 Th_{down} & : & Elev^{'} & < & -3 \\
 Th_{right} & : & Az^{'} & > & 2.5 \\
 Th_{left} & : & Az^{'} & < & -2.5 \\
 Th_{straight} & : & -1  <  & Elev^{'}~and~Az^{'} &   < 1
\end{matrix}\right.
\label{eq:thresholds}
\end{split}
\end{equation}
%
A movement is labeled with its timestamp when the gaze crosses the threshold, as shown in \Cref{fig:labeled-qvar-gaze}, while segments with more than two detected movements are discarded. For segments with exactly two movements, the algorithm checks if the combination is valid; for instance, \textit{up} and \textit{down} are ignored, but \textit{up} and \textit{left} result in the label \textit{up-left}, with the timestamp assigned to the first movement. 
%
%
An example of the final labeling according to the logger data and the derivative of the gaze data together with the synchronized QVar data is shown in \Cref{fig:labeled-qvar-gaze}. 

\section{EOG Signal Evaluation}
\label{sec:ml}


Fixed-length samples, such as one-second interval windows (equivalent to 240 data points per sample), are extracted from the dataset for analysis. 
As discussed in \Cref{sec:rel-works}, EOG signals are prone to drift over time. This behavior is also observable in the acquired QVar signals depicted in \Cref{fig:labeled-qvar-gaze}, necessitating the application of signal standardization to each window. Standardization is performed using statistics computed within the same window, as defined in \Cref{eq:standardization}, where \(\mu\) represents the mean and \(\sigma\) the standard deviation of the signal for each window.
This approach accounts for local variations within each sample window. Following standardization, filtering is applied to address the substantial noise present in the signals, as illustrated in \Cref{fig:labeled-qvar-gaze}. 
The Savitzky–Golay filter is employed, which—unlike simple moving-average filters—smooths the signal while preserving critical features, such as sharp transitions indicative of sudden movements. This is achieved by fitting a polynomial of adjustable complexity to each data window. For this study, a polynomial order of 20 is selected heuristically, as it effectively balances the removal of noise with the preservation of salient signal characteristics, as noticeable from \Cref{fig:labeled-qvar-gaze}.

\subsection{Ablation Study}

\begin{figure*}
    \centering
    \begin{subfigure}{0.98\linewidth}
        \centering
        \includegraphics[width=\linewidth]{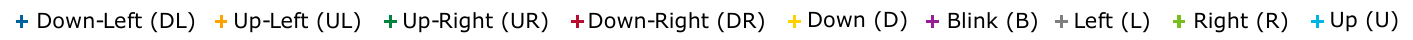}
    \end{subfigure}
    \vspace{0.5em}
    \begin{subfigure}{0.24\linewidth}
        \centering
        \includegraphics[width=\linewidth]{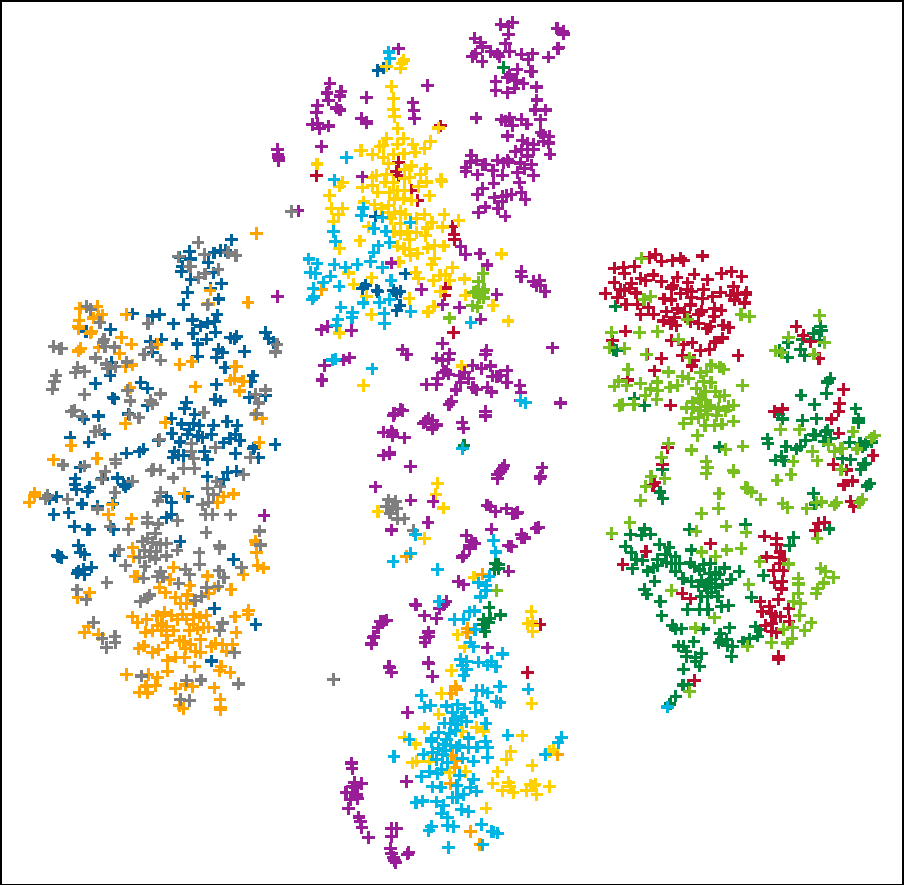}
        \caption{All ch., all classes}
        \label{fig:tsne-full}
    \end{subfigure}
    \begin{subfigure}{.24\linewidth}
        \centering
        \includegraphics[width=\linewidth]{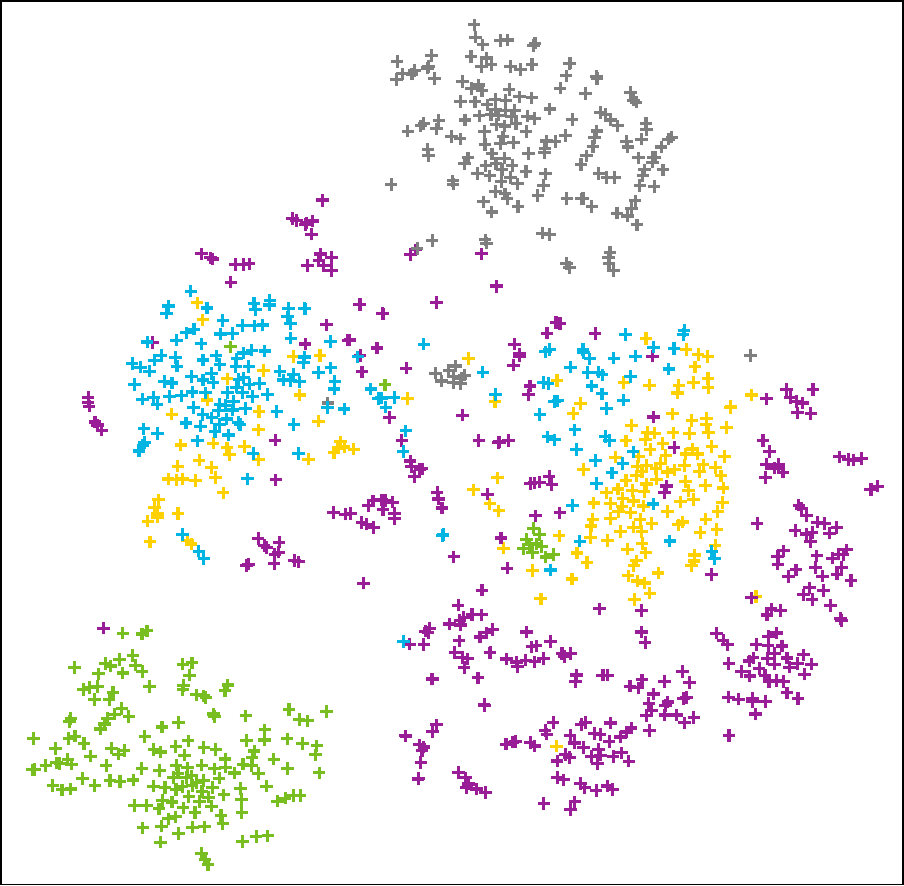}
        \caption{All ch., basic classes}
        \label{fig:tsne-basic}
    \end{subfigure}
    \begin{subfigure}{.24\linewidth}
        \centering
        \includegraphics[width=\linewidth]{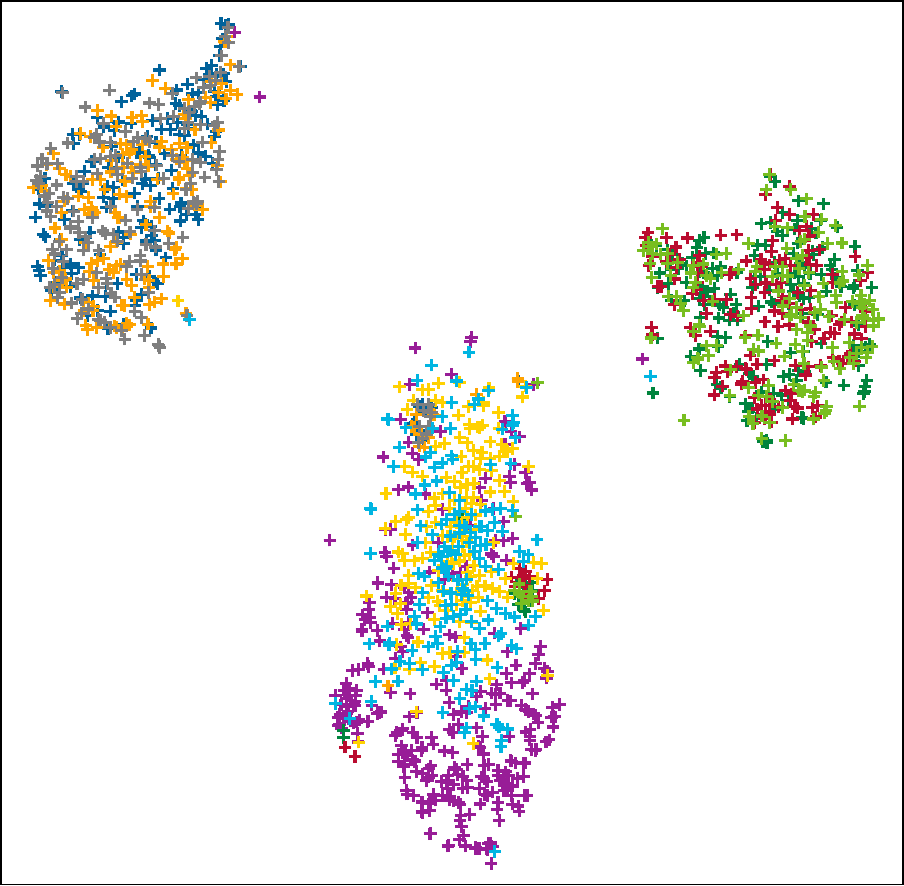}
        \caption{Contact ch., all classes}
        \label{fig:tsne-contact}
    \end{subfigure}
    \begin{subfigure}{.24\linewidth}
        \centering
        \includegraphics[width=\linewidth]{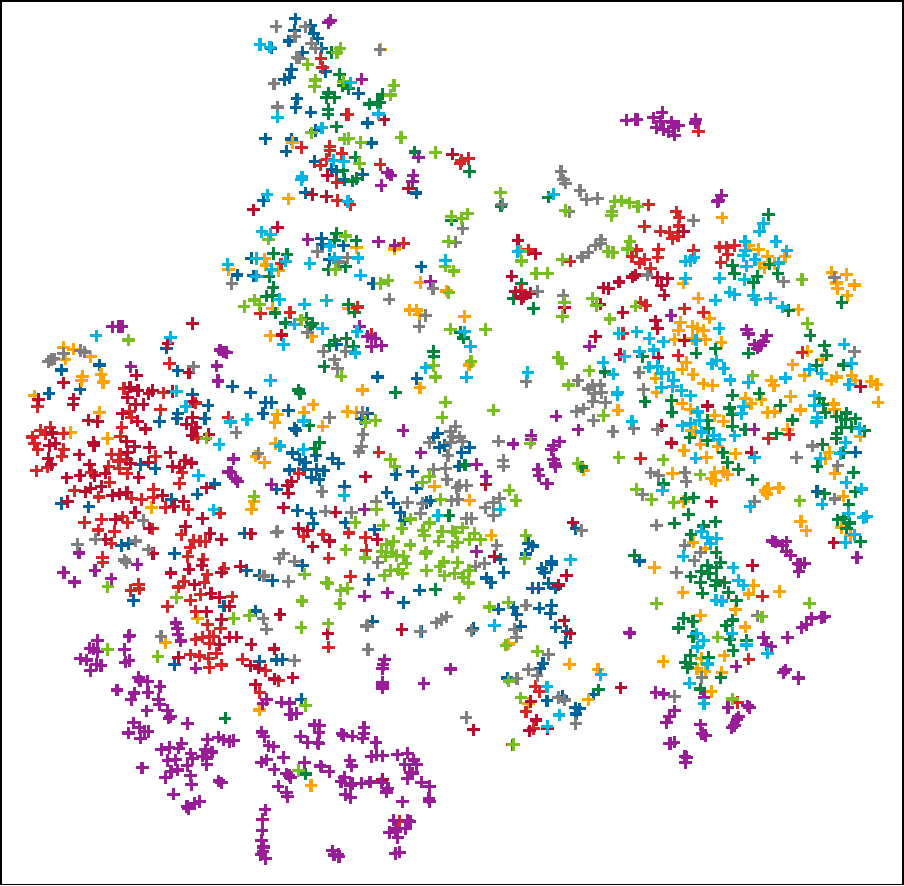}
        \caption{Contactless ch., all classes}
        \label{fig:tsne-contactless}
    \end{subfigure}
    \vspace{0.5em}
    \begin{subfigure}{0.24\linewidth}
        \centering
        \includegraphics[width=\linewidth]{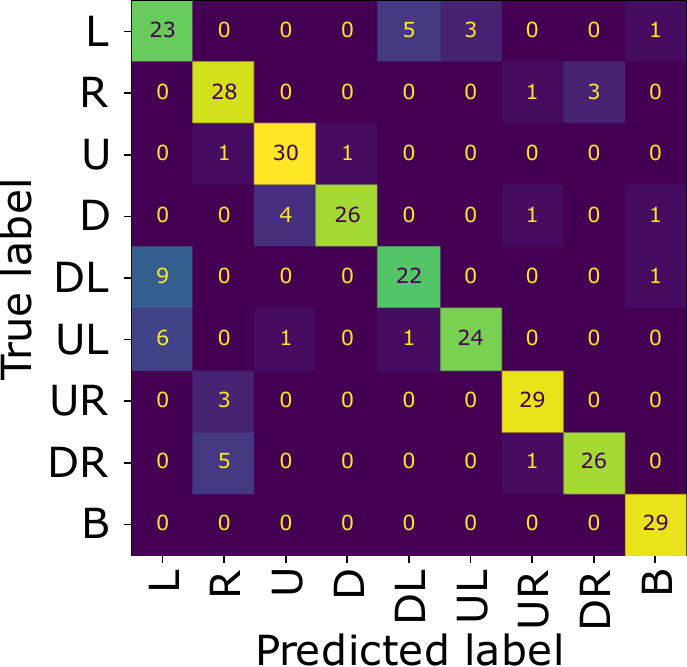}
        \caption{All ch., all classes}
        \label{fig:cm-full}
    \end{subfigure}
    \begin{subfigure}{.24\linewidth}
        \centering
        \includegraphics[width=\linewidth]{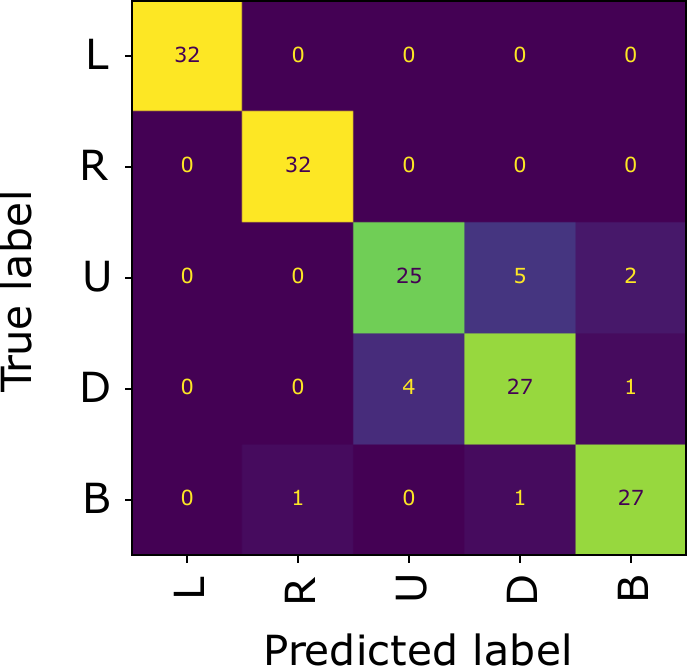}
        \caption{All ch., basic classes}
        \label{fig:cm-basic}
    \end{subfigure}
    \begin{subfigure}{.24\linewidth}
        \centering
        \includegraphics[width=\linewidth]{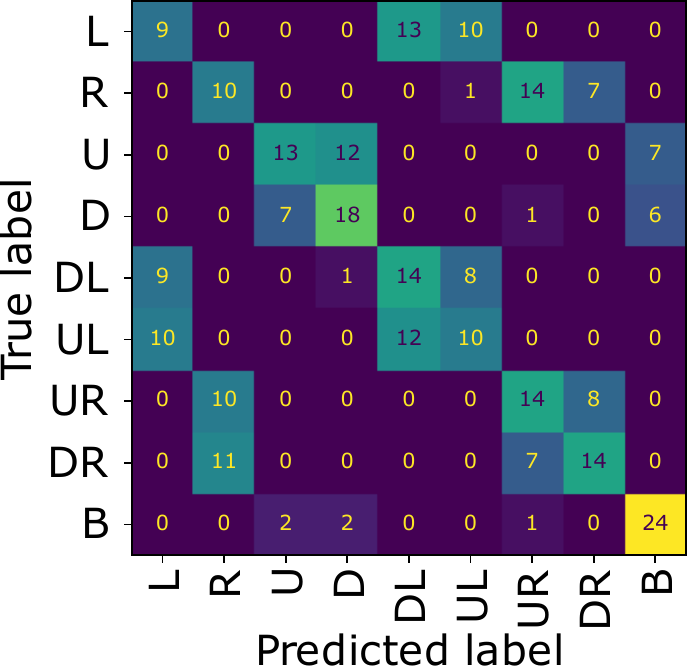}
        \caption{Contact ch., all classes}
        \label{fig:cm-contact}
    \end{subfigure}
    \begin{subfigure}{.24\linewidth}
        \centering
        \includegraphics[width=\linewidth]{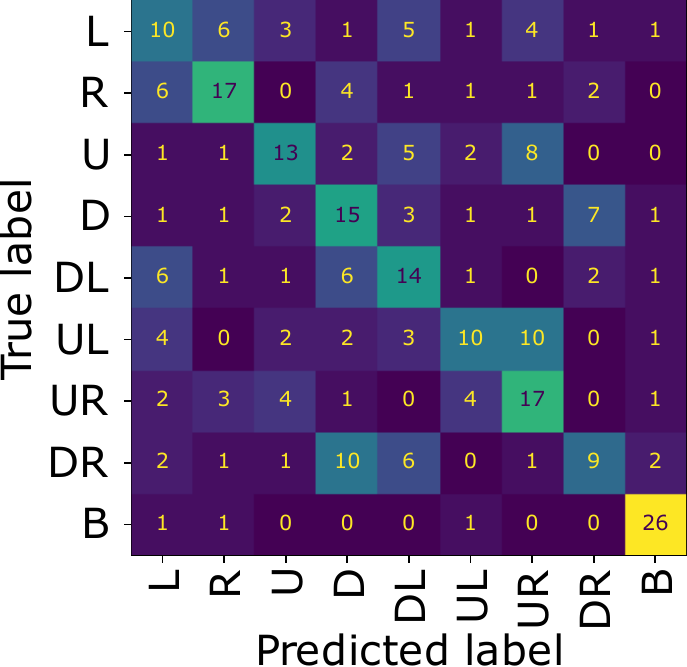}
        \caption{Contactless ch., all classes}
        \label{fig:cm-contactless}
    \end{subfigure}
    \caption{t-SNE plots (top row) and confusion matrices (bottom row) for different configurations in channels (ch.) and classes} 
    \label{fig:ablation}
\end{figure*}

An ablation study is performed to optimize the system's performance by evaluating the impact of two key parameters. The first parameter is the selection of electrode pairs, which can be \textit{all} channels, only \textit{contact} channels, or only \textit{contactless} channels. The \textit{straight} class is excluded from movement detection analysis for the ablation study, as it represents periods without movement and is therefore not relevant for detecting motion. The second parameter is the number of movements to predict, which can be either \textit{all} (9 movements) or \textit{basic} (5 movements, excluding corners). This approach helps identify the most influential factors for improving both accuracy and efficiency. Two tools are employed for this purpose: \begin{enumerate*}[label=(\roman*),font=\itshape] \item t-SNE (t-distributed Stochastic Neighbor Embedding) to provide an initial impression of the collected data, and \item a basic ML model to validate the insights gained from the t-SNE analysis and to deepen the understanding of the data. \end{enumerate*}
t-SNE is a dimensionality reduction technique commonly applied to visualize high-dimensional data. The plots are generated using a perplexity of 30 and 5000 steps. For ML, simple convolutional neural network (CNN) models with a window size of \qty{1}{\second} are trained while iterating over different parameters. The input size for the model is calculated as the number of channels multiplied by the number of QVar time points per window. The output size corresponds to the number of eye movement classes, with a maximum of nine. Each acquisition is randomly assigned to either a training set (80\% of all acquisitions) or a test set (20\% of all acquisitions). Since the model architecture is not tuned to offer the best possible performance, the results are relative, meaning that they compare to the case with the best accuracy, which is when predicting basic classes and using all the channels. The t-SNE plots and confusion matrices for four different cases are described in \Cref{fig:ablation}.

The case where all channels and all classes are used is illustrated in \Cref{fig:tsne-full,fig:cm-full}. The t-SNE plot in \Cref{fig:tsne-full} reveals three main clusters. The first cluster represents the \textit{left movements} (\textit{down-left, left, up-left}), the second corresponds to the \textit{right movements} (\textit{down-right, right, up-right}), and the third, more dispersed, represents the \textit{vertical movements} (\textit{blink, up, and down}). While the clusters are largely distinct, some overlap is observed, suggesting that the features are separable but may result in occasional classification errors within these overlapping regions. This is further supported by the confusion matrix in \Cref{fig:cm-full}, which shows minor uncertainty within the \textit{left movements}, \textit{right movements}, and \textit{vertical movements}. However, such misclassifications are infrequent, and the predictions are predominantly accurate with a relative accuracy of 91\%.

Because the highest confusion occurs within the \textit{left} and \textit{right} movement groups, the data are further evaluated by simplifying these groups into a single movement each. The results for when the number of classes is reduced to the five basic eye movements (\textit{up, down, left, right, blink}) are presented in \Cref{fig:tsne-basic,fig:cm-basic}. For t-SNE, as shown in \Cref{fig:tsne-basic}, all the clusters are easily distinguishable. There is some overlap between \textit{up} and \textit{down}, and \textit{blink} is broadly distributed. This is confirmed by the model predictions as shown in \Cref{fig:cm-basic}, where \textit{up} and \textit{down} sometimes get confused, and \textit{blink} is interpreted as other classes. However, this is not frequent and this case is the one with the highest accuracy, becoming the reference for the relative accuracy, with an accuracy of 100\%. Therefore, the assumption that reducing the number of movements to predict improves accuracy is confirmed. However, for further analysis, only the full set of classes is considered, as it provides more insight into the relevance of other parameters. The overall accuracy for the basic classes is still reported in \Cref{tab:ablation}.

A critical parameter to examine is the number of channels. The objective is to assess whether all channels are essential or if a reduced number would suffice. Minimizing the number of channels could lead to a smaller system footprint, reduced power consumption, lower processing power requirements, and an overall simplification of the system. To do so, the system is first evaluated with the contact electrodes only, and then with the contactless electrodes only.

Data for contact channels only is presented in \Cref{fig:tsne-contact,fig:cm-contact}. In the t-SNE plot, three main clusters are discernible, corresponding to the \textit{left movements} (\textit{down-left, left, up-left}), the \textit{right movements} (\textit{down-right, right, up-right}), and the \textit{vertical movements} (\textit{blink, up, down}). However, in this case, there is complete overlap within both the \textit{left} and \textit{right} movement groups, and a high degree of overlap between \textit{up} and \textit{down} with the exception of the \textit{blink} cluster, which is more dispersed. Some movements from the \textit{left} and \textit{right} movement groups are included in the \textit{vertical movements} cluster. This observation is further supported by the confusion matrix in \Cref{fig:cm-contact}, which reveals significant confusion within the \textit{left movements}, \textit{right movements}, and \textit{vertical movements}. \textit{blink} movements are more accurately detected, which aligns with the t-SNE plot. The relative accuracy is 48\%, indicating that relying solely on the contact channels is not an option.

\Cref{fig:tsne-contactless,fig:cm-contactless} presents the t-SNE plot for the data using only contactless channels. The plot reveals a dispersed distribution with some observable clusters; however, most classes—except for \textit{blinks}—show significant overlap. This observation aligns with the confusion matrix in \Cref{fig:cm-contactless}, where \textit{blinks} are well detected, but predictions for other classes are often incorrect. Although the predictions are not entirely random, the relative accuracy remains low at 50\%, indicating that relying solely on contactless electrodes is not a viable approach. All of the results described above are summarized in \Cref{tab:ablation}. When focusing on basic movements alone, the system achieves a relative accuracy of 100\% when using all channels, which serves as the reference. In this case, it is possible to consider using only contact-based or contactless channels, as the relative accuracy remains reasonably high at 77\% and 85\%, respectively. However, all movements are included, performance drops across all configurations. Using all channels achieves a relative accuracy of 91\%, while accuracy decreases to 48\% with contact-only sensing and 50\% with contactless-only sensing. The ablation study highlights that while using fewer input channels may be sufficient for simpler cases, focusing on the more complex task of detecting both basic movements and corners emphasizes the importance of using all channels.
This work primarily focuses on the case involving all movements, as it represents the most challenging scenario and enables a more meaningful comparison with the state-of-the-art.
\begin{table}
\centering
\caption{Summary of the ablation study, with relative accuracy values normalized to the maximum performance within the study.}
\label{tab:ablation}
\begin{tabular}{lcccccc}
\hline
\textbf{Channels} & \textbf{Basic} & \textbf{All} \\
\hline
\rowcolor{customgreen}
All & 100\%  & 91\% \\
Contact only & 77\% & 48\%  \\
Contactless only & 85\% & 50\%  \\
\hline
\end{tabular}
\end{table}

\section{tinyML Eye Tracking Classification Model}
\begin{figure}
    \centering
    \includegraphics[width=1\linewidth]{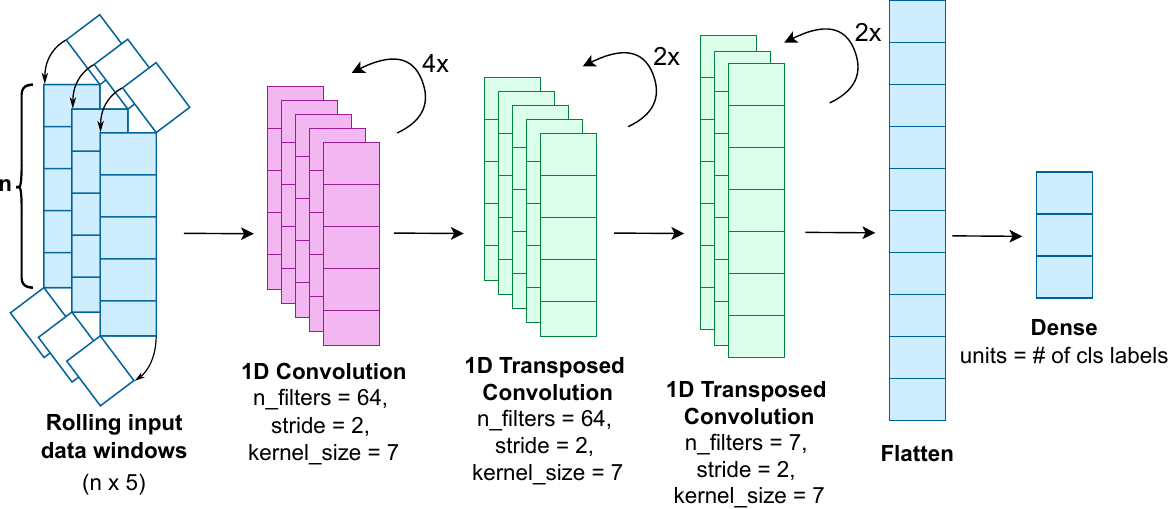}
    \caption{Architecture of the 1D-CNN model utilized in this study. Circular arrows represent repetitions of the layers when only one is shown. The variable $n$ is defined in \Cref{tab:accuracy-window-size}.}
    \label{fig:baseline-model}
\end{figure}

One of the key contributions of this paper is the design of a lightweight algorithm for eye-movement classification, optimized for execution on a low-power processing unit and leveraging the \textit{hEOG} multi-channel setup. 
As detailed in \Cref{sec:ml}, the data is first segmented into windows, which are standardized and filtered using the Savitzky-Golay filter. These pre-processed windows are then fed into a Convolutional Neural Network (CNN), whose efficient design exploits temporal information by using successive samples as input. 
In summary, the selected 1D-CNN model incorporates the following features, depicted in \Cref{fig:baseline-model}: a blend of convolutional and transposed convolutional layers, whose outputs are then flattened and passed to a dense layer. 
This layer's output is then passed through the softmax function, which generates the probability for each class label, which varies between 10 and 6. Windows of \qty{416}{\milli\second}, equivalent to 100 data points and 5 channels, are used. The first four 1D convolutions involve 64 filters with a kernel size of 7, meant to extract temporal signal dependencies from the eye movements. The  1D transposed convolutions classify the movements, featuring a kernel size of 7 with 64 and 7 channels respectively. Lastly, the flatten layer in \Cref{fig:baseline-model} calculates the class probability. These choices ensure a robust tiny model with efficient temporal feature extraction and manageable size, setting the stage for a generalized and flexible model adaptable to a multitude of users. The resulting total model dimension, trained on Tensorflow, is \qty{151447}{parameters}.

\begin{figure}[t]
    \centering
\begin{subfigure}[t]{0.24\textwidth}
    \includegraphics[width=1\linewidth]{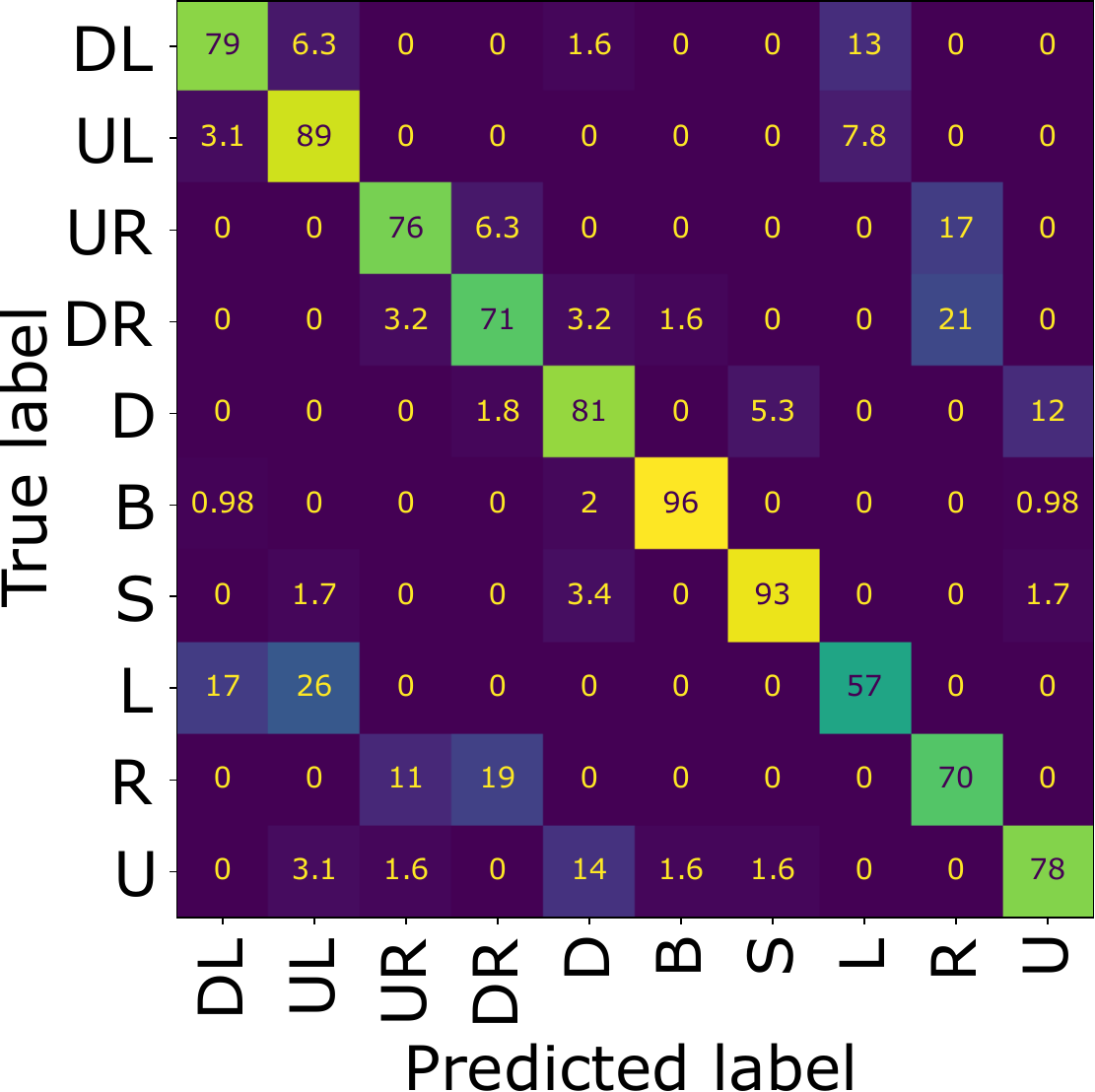}
    \caption{All 10 classes. Overall test accuracy: 81 \%.}
    \label{fig:cm_full}
\end{subfigure}
\begin{subfigure}[t]{0.24\textwidth}
    \centering
    \includegraphics[width=1\linewidth]{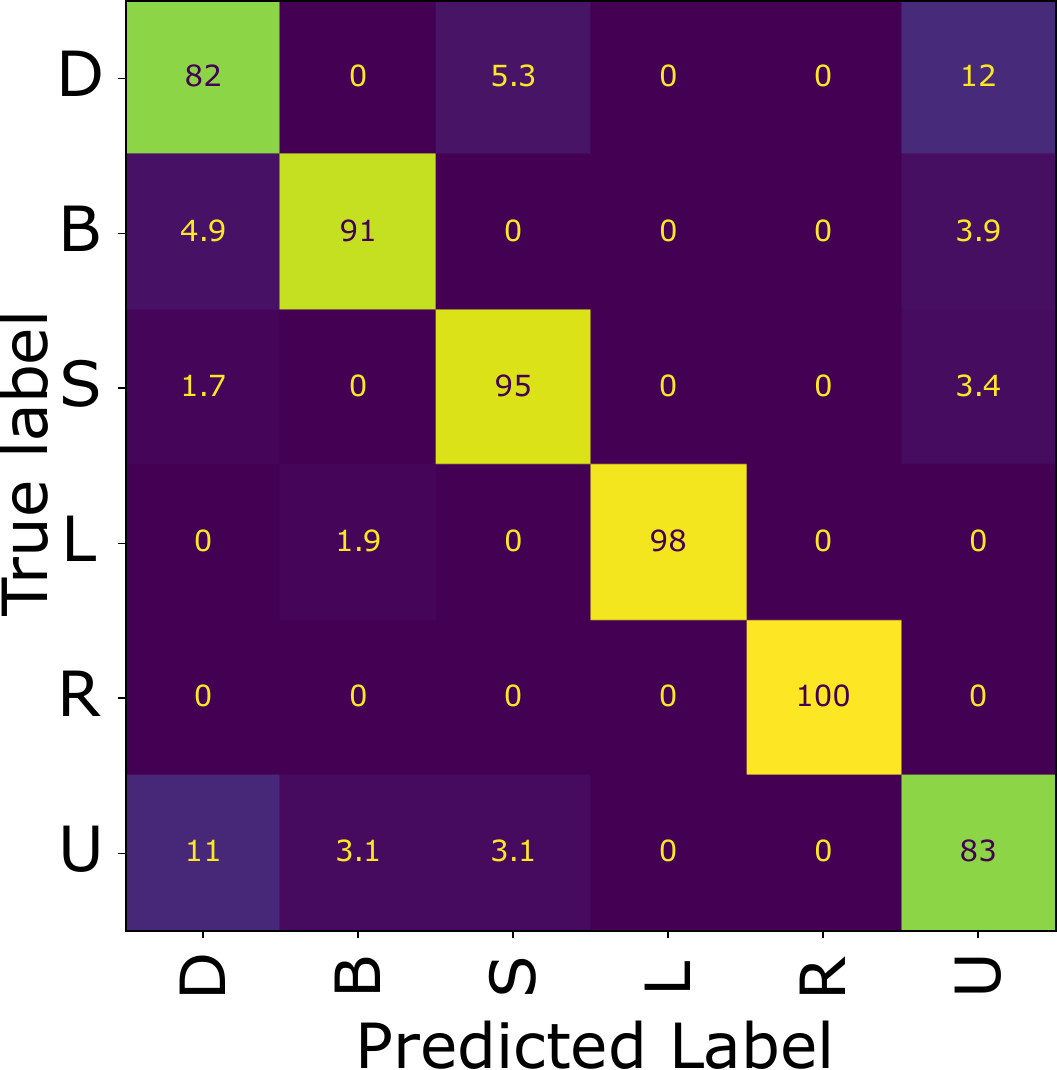}
    \caption{Basic (6) classes. Overall test accuracy: 92 \%.}
    \label{fig:cm_corners}
\end{subfigure}
\label{fig:cm}
\caption{Normalized confusion matrices of the two full-precision models with a window size of \qty{416}{\milli\second}.}
\end{figure}

After parameter fine-tuning and training, the classification performance of the tinyML model is presented in \Cref{fig:cm_full}, demonstrating an overall test accuracy of 81\%, evaluated on subjects not encountered during training. The model effectively distinguishes different movements, although the main sources of confusion are between the corner labels and the corresponding horizontal movements. To further evaluate this observation, the classification performance of the same model trained and tested on all basic movements is presented in \Cref{fig:cm_corners}. This yields an accuracy of 92\%.

\subsection{Optimal Input Window Size}

To enhance the system, the tiny CNN model is evaluated across various window sizes, representing the input layer length, with a comprehensive analysis provided in \Cref{tab:accuracy-window-size}.
A common misconception is that window size is directly linked to latency. In reality, latency refers to the time between the occurrence of a movement and the end of the window, meaning a longer window does not necessarily result in higher latency. The goal of reducing the window size, therefore, is not to decrease latency but to enable the detection of more movements within a given time frame, while reducing memory usage and computational requirements.
To have an accurate evaluation of the proposed model, this paper introduces a novel ground-truth method using the PL Neon camera-based eye tracking. 
\begin{table}[t]
   \centering
      \caption{Absolute accuracy on the test set for the baseline model with 10 classes, evaluated across various window sizes. The window length in number of data points, denoted as $n$, represents the input size of the CNN.}
    \begin{tabular}{ccc}
        \hline
        \multicolumn{2}{c}{\textbf{Window size}} & \textbf{1D-CNN model} \\
        \cline{1-2}
        \textbf{Data points (\textit{n})} & \textbf{Milliseconds} & \textbf{absolute accuracy} \\
        \hline
         240 & 1000 & 84\% \\
         200 & 833 & 85\% \\
         150 & 625 & 83\% \\
         125 & 520 & 82\% \\
         \rowcolor{customgreen}
         100 & 416 & 81\% \\
         75 & 312 & 73\% \\
         50 & 208 & 64\% \\
         25 & 104 & 44\% \\
        \hline
        \\
    
    \end{tabular}
   \label{tab:accuracy-window-size}
\end{table}
%


As detailed in \Cref{tab:accuracy-window-size}, the accuracy remains stable across a broad range of window sizes, specifically between \qty{1}{\second} and \qty{416}{\milli\second}, with accuracies ranging from 81\% to 85\%. However, as the window size decreases, the accuracy drops, reaching 73\% for \qty{312}{\milli\second} and 64\% for \qty{208}{\milli\second}.
This allows the system to detect up to five movements per second, which exceeds the typical frequency of human eye movements, such as saccades, which occur 3 to 4 times per second~\cite{brown_iscev_2006}.
Additionally, the accuracy is evaluated using the logger labels, which consistently perform worse than the eye-tracker-based labels. For instance, the logger labels are $4.6 \times$ less accurate than the eye-tracker labels in predicting rapid movements at this particular window size. This demonstrates the superiority of the presented novel eye tracking labeling method.
Based on these results, the optimal window size to decrease the RAM usage while maintaining real-time efficiency is \qty{416}{\milli\second}, or 100 data points, using the PL Neon eye-tracker ground truth. This solution offers an optimal balance between accuracy and the number of movements detectable within a given timeframe.

\subsection{Optimized Real-Time Inference Onboard Smart Glasses}
During the training phase performed offline, the entire dataset is processed with the start and end of each movement clearly labeled. This approach assumes that every input window contains a detectable movement, so the model is applied directly to these predefined segments for classification. In this scenario, each window is known to contain either a movement or a specific label, such as "blink", simplifying the model’s task. The challenge is to classify which specific movement occurred within the window. Moreover, some windows might not have a movement at all. To avoid the model predicting a random class when this condition happens, a specific  \textit{straight} class is introduced. 

Real-time inference, however, introduces a different challenge. In live data streams, it is not clear whether or when a movement occurs. Therefore, the data must be processed using rolling windows, where overlapping segments are continuously extracted from the data stream. Each window may or may not contain a movement, and the system must make a decision in real-time to determine whether a movement has occurred and, if so, what type. 

\subsection{Prediction Latency}
\label{sec:latency}
The rolling window approach is utilized to estimate the model's latency, not considering the processing time. Data from three random acquisitions are segmented into rolling windows of \qty{416}{\milli\second} (100 data points), with a stride of \qty{8}{\milli\second} (two data points) between consecutive windows to maintain temporal continuity and ensure that no movements occurring near window boundaries are missed. 

\begin{figure}
    \centering
    \includegraphics[width=.9\linewidth]{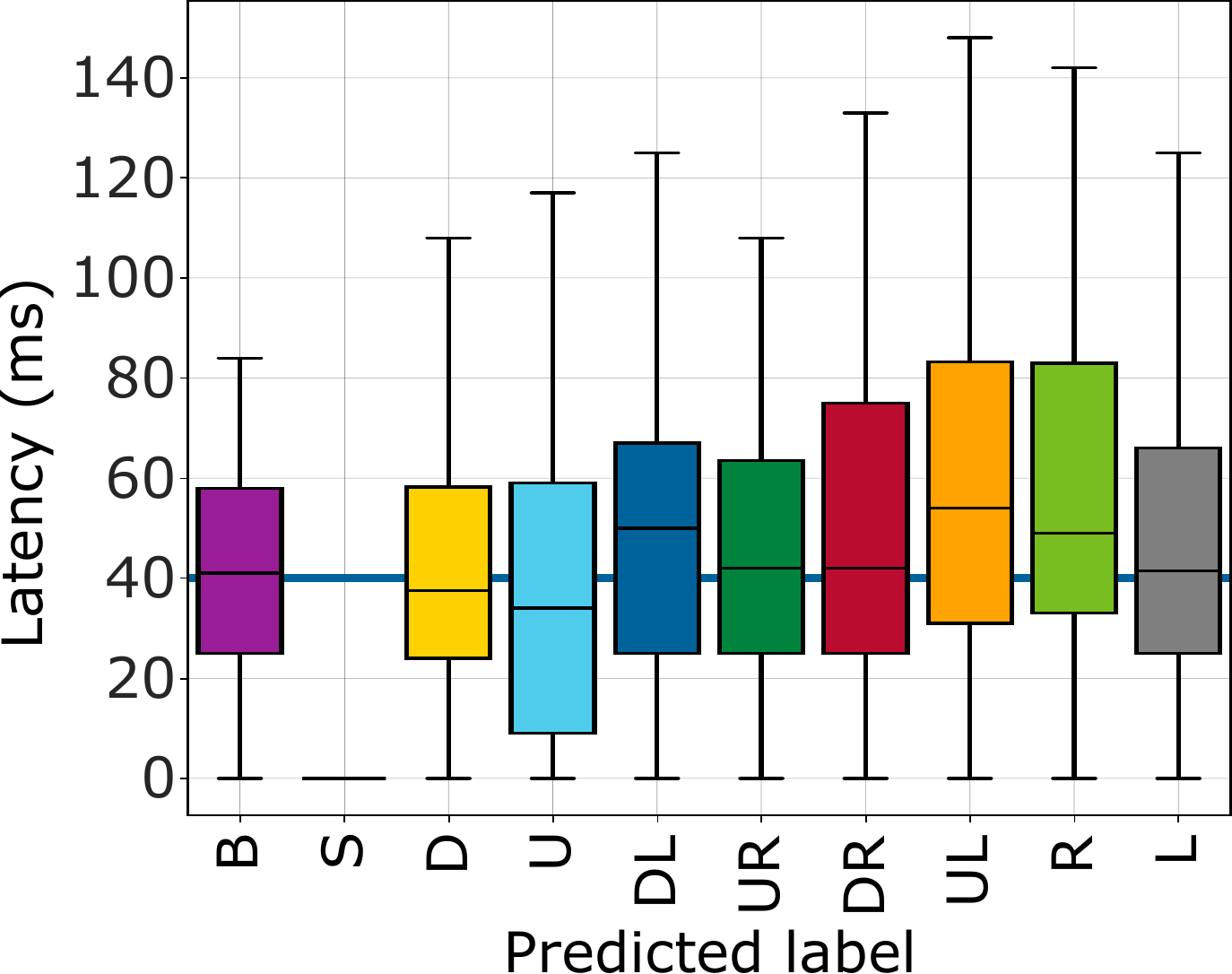}
    \caption{Latency from the start of the movement to the model's correct prediction of the movement. The blue line represents the average duration of a human eye movement (\qty{40}{\milli\second}).}
    \label{fig:movement-latency}
\end{figure}

For each given movement occurring at timestamp $t_m$, the first rolling window that is passed through the model starts at timestamp $t_m$ - \qty{416}{\milli\second}. This ensures that the start of the movement is located at the last timestamp of the first window. Each subsequent rolling window $i$ starts at timestamp $t_m$ - \qty{416}{\milli\second} + (8 * $i$)~ms, until the last rolling window for a given movement, which starts at timestamp $t_m$. This ensures a fair performance evaluation of the model whilst simulating live conditions. This approach also requires the use of precise movement labeling in time, which further emphasizes the importance of a reliable ground truth, such as the one used in this study. 
The latency for a movement is calculated if the movement is correctly predicted, and is equal to the difference between the timestamp of the end of the correctly predicted window, and the timestamp of the movement in the ground truth, i.e., at the beginning of the movement.
In this study, each movement of the test set is passed through the model using this method. During this process, only 1\% of the test samples are never correctly predicted. This confirms the relevance of the model as well as the window size. 
The measured latency for each class is represented as a box plot in \Cref{fig:movement-latency}. 

The \textit{straight} class is always correctly predicted with zero latency being the default model response. 
Analyzing the ground truth data and the literature~\cite{brown_iscev_2006} reveals that eye movements typically take \qty{40}{\milli\second}. The median latency of the presented model in \Cref{fig:movement-latency} is also \qty{40}{\milli\second}, indicating that half of the movements are accurately predicted before their completion. 

Building on this observation, calculating the latency from the end of the movement rather than its beginning demonstrates that \qty{40}{\milli\second} is sufficient to correctly predict over 75\% of movements, while \qty{60}{\milli\second} covers 90\%.
The ground truth data also shows that smaller movements require less time than larger ones. This is reflected in \Cref{fig:movement-latency}, where \textit{up} and \textit{down} movements —shorter due to the position of the dot during data acquisition— exhibit lower latencies compared to other eye movements.
These findings highlight the model's ability to accurately detect eye movements within a narrow time frame relative to the movement itself. Moreover, the novel non-invasive hEOG solution can accurately detect eye movements in real-time with a latency imperceptible to users, enhancing their overall experience.

\subsection{GAP9 Deployment for Real-Time Eye Tracking}
The trained model is optimized, quantized, and executed on the GAP9 MCU embedded in the tinyML VitalPack, with a fixed train/test split yielding a full-precision test accuracy of 81\%, as seen in \Cref{tab:gap9_deply}. The original model, in \textit{float32}, is quantized at different bit-precisions to find the best compromise between execution time, energy per inference, memory footprint, and accuracy. The deployment and quantization are performed using the NNTool from Greenwaves Technologies. \Cref{tab:gap9_deply} shows the quantization scheme results, exploiting the GAP9 CL at \qty{370}{\mega\hertz} with an average measured power consumption of \qty{153}{\milli\watt} during inference. This power consumption includes external peripherals such as memory and power supply. In \Cref{tab:gap9_deply}, while the model size scales (almost) linearly with the applied quantization scheme, the accuracy remains constant down to \textit{4-bit}. On the other side, the execution time, and consequently the energy per inference, are drastically reduced from \textit{float16} with $\sim$\qty{2}{\milli\second} to \textit{8-bit} with $\sim$\qty{330}{\micro\second}. However, the same behavior is not noticed between 8 and \textit{4-bit}, where the execution time decreases by 9\%. With a \textit{2-bit} quantization, the model does not operate anymore, dropping the accuracy down to 11\%.

To enhance inference speed and energy efficiency on the presented low-power wearable platform, the \textit{4-bit} model is employed for field experimental results, with a model size of only \qty{79}{\kilo B}, which fits entirely into the L1 cache of GAP9. This quantization results in a test accuracy of 80\%, comparable to the full-precision model’s accuracy, as shown in \Cref{tab:gap9_deply}. The \textit{4-bit} quantized model offers the best trade-off between accuracy and cycles per inference, requiring \qty{111000}{cycles} and an equivalent execution time of \qty{301}{\micro\second} per inference. This efficient memory usage and reduced power consumption (\qty{46}{\micro\joule}) make the \textit{4-bit} quantized model suitable for real-time and low latency operations on resource-constrained devices like low-power eye tracking.
Moreover, the execution time of  \qty{301}{\micro\second} marginally affects the overall system latency investigated in \Cref{fig:movement-latency}, increasing the eye classification delay by a mere 0.7\%.

\begin{table}[t]
    \centering
    \caption{Summary of different quantization schemes deployed on GAP9. Execution time and energy consider one inference on the GAP9 cluster at \qty{370}{\mega\hertz} and 10 classes.}
    \begin{tabular}{lccccc}
    \hline
            \begin{tabular}[c]{@{}c@{}} \textbf{Quant.}  \end{tabular} & \begin{tabular}[c]{@{}c@{}} \textbf{Model} \\ \textbf{size {[}kB{]}}\end{tabular} & \begin{tabular}[c]{@{}c@{}}\textbf{Cycles} \\ $\times\mathbf{10^3}$ \end{tabular} & \begin{tabular}[c]{@{}c@{}}\textbf{Accuracy} \\ \textbf{[\%]}\end{tabular}& \begin{tabular}[c]{@{}c@{}}\textbf{Exec.}\\ \textbf{time {[}\(\mu s\){]}} \end{tabular} & \begin{tabular}[c]{@{}c@{}}\textbf{Energy}$^\star$ \\\textbf{ {[}\(\mu J\){]}} \end{tabular} \\
    \hline
    \textit{float32} & 592                  & -      & 81        & -                  &     -                          \\
    \textit{float16} & 303             & 746 & 81          & 2017               & 309                           \\
    \textit{8-bit}   & 155             & 122 & 81          & 330                & 50                            \\
    \rowcolor{customgreen}
    \textit{4-bit}   & 79              & 111 & 80         & 301                & 46                            \\
    \textit{2-bit}   & 41              & 107 & 11         & 289                & 44                            \\
    \hline
    \multicolumn{6}{l}{$^\star$ Measured average power \qty{153}{\milli\watt}}
    \end{tabular}

    \label{tab:gap9_deply}
\end{table}
\section{Power Consumption and Battery Lifetime}
Measurements are conducted on the prototype with field experiments, featuring fully onboard sensing and computation, to assess 
\begin{enumerate*}[label=(\roman*),font=\itshape]
\item the power consumption during sensor sampling, and 
\item the energy for a GAP9 operation while performing inference for eye movement prediction.
\end{enumerate*}
\subsection{Sensor Data Sampling}
Power measurements are performed on the VitalCore with the QVar VitalPack attached using the \textit{Keysight NC065C DC Power Analyzer}. When no sensors are active, the power baseline power consumption is \qty{4.0}{\milli\watt}. Power is measured for varying numbers of active sensors, resulting in an average consumption of \qty{0.75}{\milli\watt} per sensor. This value accounts for the MCU operating in active mode and sampling the ST1VAFE3BX via SPI. Therefore, the power consumption for data acquisition of ElectraSight is \qty{7.75}{\milli\watt}. 

\subsection{Onboard CNN Inference}

As detailed in \Cref{sec:ml}, the energy required for a single eye movement prediction using the \textit{4-bit} quantized model is \qty{46}{\micro\joule}. Live inference, as explained in \Cref{sec:ml}, requires overlapping windows. A higher overlap increases the number of predictions and, consequently, power consumption. This solution provides flexibility to balance accuracy and power consumption based on application requirements: a high overlap prioritizes accuracy and low latency, while a reduced overlap minimizes power usage. \Cref{tab:inference-consumption} summarizes these trade-offs.

\begin{table}[t]
    \centering
    \caption{Power consumption and battery lifetime estimation for different overlaps in sliding windows for real-time eye-movement prediction.}
    \begin{tabular}{cccc}
    \hline
    \textbf{Overlap (\%)} & \textbf{GAP9 (mW)} & \textbf{Sliding time (ms)} & \textbf{Battery (days)} \\ 
    \hline
    99 & 11.04 & 4.20 & 1.4        \\
    98 & 5.52  & 8.40 & 2.0        \\
    96 & 2.76  & 17.0 & 2.6        \\
    \rowcolor{customgreen}
    90 & 1.10  & 42.0 & 3.0        \\
    52 & 0.23  & 191  & 3.4        \\ 
    \hline
    \end{tabular}
    \label{tab:inference-consumption}
\end{table}

An overlap of 90\% presents a good compromise between power consumption and sliding time, consuming only \qty{1.10}{\milli\watt} for processing. Notably, the sliding time of this configuration aligns with the human eye average moving time discussed in \Cref{sec:latency}. 

In the selected working mode, the whole system uses \qty{7.75}{\milli\watt} for data acquisition and \qty{1.10}{\milli\watt} for computing, for a total system average power consumption of \qty{8.85}{\milli\watt}. Therefore, ElectraSight can operate for three days on the \qty{175}{\milli\ampere\hour} embedded battery, using the configuration shown in \Cref{fig:inference-system-description}.


\section{Conclusions}
\label{sec:discussion}
\label{sec:conclusion}

ElectraSight demonstrates high accuracy in detecting various eye movements using a plug-and-play non-invasive setup. The system’s design process encompasses sensor characterization and the evaluation of optimal electrode placement and impedance, leading to a fully integrated low-power solution embedded into a regular glasses frame. 
By incorporating non-contact sensors, ElectraSight sets itself apart from traditional fully contact-based EOG systems, offering a \textit{wear-and-forget} design that is calibration-free, non-stigmatizing, and user-agnostic—effectively addressing the limitations of intrusive and uncomfortable alternatives.


The system achieves a classification accuracy of 92\% for basic movements (6 classes). When corner movements are included (10 classes), accuracy decreases to 81\%. Ablation studies underscore the value of the hybrid sensing approach: removing the non-contact channels reduces relative accuracy from 91\% to 50\% for all movements, highlighting their critical contribution. Additionally, classification latency remains below \qty{60}{\milli\second} for most movements, supporting real-time applications. Notably, half of the movements can be predicted even before they conclude.

ElectraSight also outperforms competing systems in processing speed by orders of magnitude, with a computation time of just \qty{301}{\micro\second}. 
Compared to state-of-the-art solutions, it delivers competitive accuracy while excelling in power efficiency and user comfort. For instance,\cite{das_eog_2024} achieves 95\% accuracy across six categories using wet-electrode EOG, but its high power consumption (\qty{1446}{\milli\watt}) and invasive design preclude integration into glasses frames. Similarly, while\cite{frey_gapses_2024} embeds its system into glasses and achieves 95\% accuracy, it is constrained to detecting one movement every four seconds and requires user-specific calibration and re-training. In contrast, ElectraSight enables real-time, low-latency classification without user-specific adjustments.
Unlike solutions such as~\cite{shi_eye_2023}, which achieve 97\% accuracy on nine categories using QVar sensors but focus on gaze estimation, ElectraSight prioritizes practical movement classification for real-time applications. Moreover, its use of low-power, off-the-shelf components offers a cost-effective and robust alternative. Compared to~\cite{li_gazetrak_2024}, which employs acoustic sensing, ElectraSight is less power-intensive and unaffected by external sound interference, ensuring greater reliability for wearable applications.

With onboard real-time processing, low power consumption (\qty{8.85}{\milli\watt}), and a three-day battery life, ElectraSight emerges as a versatile solution for continuous-use wearable applications. 

Future developments will aim to enhance the movement prediction algorithm and explore the potential of full QVar-based eye tracking.





\bibliography{references}
\bibliographystyle{IEEEtran}

\end{document}